\documentclass[energies,article,accept,moreauthors,pdftex]{Definitions/mdpi}

\usepackage[normalem]{ulem} 
\usepackage{amsfonts} 
\usepackage{wasysym} 
\usepackage{mathcomp} 
\usepackage{CJKutf8} 
\usepackage{pifont} 
\usepackage{bm} 
\usepackage{bbm} 
\graphicspath{{./Definitions/}} 

\makeatletter
\def\T@n@@nc@d@ngM@cr@M@d{}
\def\LY@n@@nc@d@ngM@cr@M@d{}
\makeatother

\let\orignewcommand\newcommand  
\let\newcommand\providecommand  
\usepackage{verse}
\let\newcommand\orignewcommand  

\newsavebox\foobox

\renewcommand{\dagger}{\mathchar"2279}
\renewcommand{\ddagger}{\mathchar"227A}

\newcommand{\mmathit}[1]{
  \ifthenelse{\equal{#1}{\ln}}{\mathit{ln}}{
    \ifthenelse{\equal{#1}{\max}}{\mathit{max}}{\mathit{#1}}
  }
}
\makeatother
\robustify{\footnote} 
  
\DeclareUnicodeCharacter{1E45}{\.{n}}
\DeclareUnicodeCharacter{1E41}{\.{m}}
\DeclareUnicodeCharacter{2003}{\quad}
\DeclareUnicodeCharacter{2009}{\thinspace}
\DeclareUnicodeCharacter{2002}{\enspace{}}
\DeclareUnicodeCharacter{2005}{\thinspace}
\DeclareUnicodeCharacter{0263}{\textipa{G}}
\DeclareUnicodeCharacter{A0}{~}
\DeclareUnicodeCharacter{2460}{\textcircled{\scriptsize{1}}}
\DeclareUnicodeCharacter{2461}{\textcircled{\scriptsize{2}}}
\DeclareUnicodeCharacter{2462}{\textcircled{\scriptsize{3}}}
\DeclareUnicodeCharacter{2463}{\textcircled{\scriptsize{4}}}
\DeclareUnicodeCharacter{2464}{\textcircled{\scriptsize{5}}}
\DeclareUnicodeCharacter{2465}{\textcircled{\scriptsize{6}}}
\DeclareUnicodeCharacter{2466}{\textcircled{\scriptsize{7}}}
\DeclareUnicodeCharacter{2467}{\textcircled{\scriptsize{8}}}
\DeclareUnicodeCharacter{2468}{\textcircled{\scriptsize{9}}}
\DeclareUnicodeCharacter{2070}{\textsuperscript{0}}
\DeclareUnicodeCharacter{2074}{\textsuperscript{4}}
\DeclareUnicodeCharacter{2075}{\textsuperscript{5}}
\DeclareUnicodeCharacter{2076}{\textsuperscript{6}}
\DeclareUnicodeCharacter{2077}{\textsuperscript{7}}
\DeclareUnicodeCharacter{2078}{\textsuperscript{8}}
\DeclareUnicodeCharacter{2079}{\textsuperscript{9}}
\DeclareUnicodeCharacter{02C2}{<}
\DeclareUnicodeCharacter{2033}{\relax\ifmmode '' \else $''$\fi}
\DeclareUnicodeCharacter{2034}{\relax\ifmmode ''' \else $'''$\fi}
\DeclareUnicodeCharacter{2026}{\relax\ifmmode … \else $\ldots$\fi}
\DeclareUnicodeCharacter{0229}{\c{e}}
\DeclareUnicodeCharacter{016F}{\r{u}}
\DeclareUnicodeCharacter{127}{\relax\ifmmode\rm\hbar\else $\rm\hbar$\fi}
\DeclareUnicodeCharacter{3AC}{\relax\ifmmode\acute{\alpha}\else $\acute{\alpha}$\fi}
\DeclareUnicodeCharacter{3AD}{\relax\ifmmode\acute{\varepsilon}\else $\acute{\varepsilon}$\fi}
\DeclareUnicodeCharacter{3AE}{\relax\ifmmode\acute{\eta}\else $\acute{\eta}$\fi}
\DeclareUnicodeCharacter{3AF}{\relax\ifmmode\acute{\iota}\else $\acute{\iota}$\fi}
\DeclareUnicodeCharacter{3CC}{\relax\ifmmode\acute{o}\else $\acute{o}$\fi}
\DeclareUnicodeCharacter{3CD}{\relax\ifmmode\acute{\upsilon}\else $\acute{\upsilon}$\fi}
\DeclareUnicodeCharacter{3CE}{\relax\ifmmode\acute{\omega}\else $\acute{\omega}$\fi}
\DeclareUnicodeCharacter{391}{A}
\DeclareUnicodeCharacter{392}{B}
\DeclareUnicodeCharacter{395}{E}
\DeclareUnicodeCharacter{396}{Z}
\DeclareUnicodeCharacter{397}{H}
\DeclareUnicodeCharacter{399}{I}
\DeclareUnicodeCharacter{39A}{K}
\DeclareUnicodeCharacter{39C}{M}
\DeclareUnicodeCharacter{39D}{N}
\DeclareUnicodeCharacter{39F}{O}
\DeclareUnicodeCharacter{3A1}{P}
\DeclareUnicodeCharacter{3A4}{T}
\DeclareUnicodeCharacter{3A7}{X}

\DeclareUnicodeCharacter{27E6}{\relax\ifmmode \llbracket \else $\llbracket$\fi}
\DeclareUnicodeCharacter{27E7}{\relax\ifmmode \rrbracket \else $\rrbracket$\fi}

\DeclareUnicodeCharacter{1D434}{\relax\ifmmode A \else $A$\fi}
\DeclareUnicodeCharacter{1D435}{\relax\ifmmode B \else $B$\fi}
\DeclareUnicodeCharacter{1D436}{\relax\ifmmode C \else $C$\fi}
\DeclareUnicodeCharacter{1D437}{\relax\ifmmode D \else $D$\fi}
\DeclareUnicodeCharacter{1D438}{\relax\ifmmode E \else $E$\fi}
\DeclareUnicodeCharacter{1D439}{\relax\ifmmode F \else $F$\fi}
\DeclareUnicodeCharacter{1D43A}{\relax\ifmmode G \else $G$\fi}
\DeclareUnicodeCharacter{1D43B}{\relax\ifmmode H \else $H$\fi}
\DeclareUnicodeCharacter{1D43C}{\relax\ifmmode I \else $I$\fi}
\DeclareUnicodeCharacter{1D43D}{\relax\ifmmode J \else $J$\fi}
\DeclareUnicodeCharacter{1D43E}{\relax\ifmmode K \else $K$\fi}
\DeclareUnicodeCharacter{1D43F}{\relax\ifmmode L \else $L$\fi}
\DeclareUnicodeCharacter{1D440}{\relax\ifmmode M \else $M$\fi}
\DeclareUnicodeCharacter{1D441}{\relax\ifmmode N \else $N$\fi}
\DeclareUnicodeCharacter{1D442}{\relax\ifmmode O \else $O$\fi}
\DeclareUnicodeCharacter{1D443}{\relax\ifmmode P \else $P$\fi}
\DeclareUnicodeCharacter{1D444}{\relax\ifmmode Q \else $Q$\fi}
\DeclareUnicodeCharacter{1D445}{\relax\ifmmode R \else $R$\fi}
\DeclareUnicodeCharacter{1D446}{\relax\ifmmode S \else $S$\fi}
\DeclareUnicodeCharacter{1D447}{\relax\ifmmode T \else $T$\fi}
\DeclareUnicodeCharacter{1D448}{\relax\ifmmode U \else $U$\fi}
\DeclareUnicodeCharacter{1D449}{\relax\ifmmode V \else $V$\fi}
\DeclareUnicodeCharacter{1D44A}{\relax\ifmmode W \else $W$\fi}
\DeclareUnicodeCharacter{1D44B}{\relax\ifmmode X \else $X$\fi}
\DeclareUnicodeCharacter{1D44C}{\relax\ifmmode Y \else $Y$\fi}
\DeclareUnicodeCharacter{1D44D}{\relax\ifmmode Z \else $Z$\fi}
\DeclareUnicodeCharacter{1D44E}{\relax\ifmmode a \else $a$\fi}
\DeclareUnicodeCharacter{1D44F}{\relax\ifmmode b \else $b$\fi}
\DeclareUnicodeCharacter{1D450}{\relax\ifmmode c \else $c$\fi}
\DeclareUnicodeCharacter{1D451}{\relax\ifmmode d \else $d$\fi}
\DeclareUnicodeCharacter{1D452}{\relax\ifmmode e \else $e$\fi}
\DeclareUnicodeCharacter{1D453}{\relax\ifmmode f \else $f$\fi}
\DeclareUnicodeCharacter{1D454}{\relax\ifmmode g \else $g$\fi}
\DeclareUnicodeCharacter{1D456}{\relax\ifmmode i \else $i$\fi}
\DeclareUnicodeCharacter{1D457}{\relax\ifmmode j \else $j$\fi}
\DeclareUnicodeCharacter{1D458}{\relax\ifmmode k \else $k$\fi}
\DeclareUnicodeCharacter{1D459}{\relax\ifmmode l \else $l$\fi}
\DeclareUnicodeCharacter{1D45A}{\relax\ifmmode m \else $m$\fi}
\DeclareUnicodeCharacter{1D45B}{\relax\ifmmode n \else $n$\fi}
\DeclareUnicodeCharacter{1D45C}{\relax\ifmmode o \else $o$\fi}
\DeclareUnicodeCharacter{1D45D}{\relax\ifmmode p \else $p$\fi}
\DeclareUnicodeCharacter{1D45E}{\relax\ifmmode q \else $q$\fi}
\DeclareUnicodeCharacter{1D45F}{\relax\ifmmode r \else $r$\fi}
\DeclareUnicodeCharacter{1D460}{\relax\ifmmode s \else $s$\fi}
\DeclareUnicodeCharacter{1D461}{\relax\ifmmode t \else $t$\fi}
\DeclareUnicodeCharacter{1D462}{\relax\ifmmode u \else $u$\fi}
\DeclareUnicodeCharacter{1D463}{\relax\ifmmode v \else $v$\fi}
\DeclareUnicodeCharacter{1D464}{\relax\ifmmode w \else $w$\fi}
\DeclareUnicodeCharacter{1D465}{\relax\ifmmode x \else $x$\fi}
\DeclareUnicodeCharacter{1D466}{\relax\ifmmode y \else $y$\fi}
\DeclareUnicodeCharacter{1D467}{\relax\ifmmode z \else $z$\fi}

\DeclareUnicodeCharacter{1E67}{\.{\v s}}
\DeclareUnicodeCharacter{1E11}{\relax\ifmmode \c{d} \else $\c{d}$\fi}
\DeclareUnicodeCharacter{1ECB}{\relax\ifmmode \d{i} \else $\d{i}$\fi}
\DeclareUnicodeCharacter{1D8D}{\relax\ifmmode \textlhookx \else $\textlhookx$\fi}
\DeclareUnicodeCharacter{104}{\relax\ifmmode \k{A} \else $\k{A}$\fi}
\DeclareUnicodeCharacter{211E}{\relax\ifmmode \textrecipe \else $\textrecipe$\fi}
\DeclareUnicodeCharacter{29D}{\relax\ifmmode \textctj \else $\textctj$\fi}

\DeclareUnicodeCharacter{1E2E}{\'{\"I}}
\DeclareUnicodeCharacter{23F}{\textrts}

\DeclareUnicodeCharacter{2C73}{\varw}

\DeclareUnicodeCharacter{2127}{\mho}

\DeclareUnicodeCharacter{28C}{\textturnv}
\DeclareUnicodeCharacter{252}{\textturnscripta}
\DeclareUnicodeCharacter{259}{\schwa}
\DeclareUnicodeCharacter{25B}{\m{e}}
\DeclareUnicodeCharacter{266}{\m{h}}
\DeclareUnicodeCharacter{127}{\B{h}}
\DeclareUnicodeCharacter{27E}{\textfishhookr}
\DeclareUnicodeCharacter{281}{\textinvscr}

\firstpage{1}
  \articlenumber{3769}
\pubvolume{18}
\issuenum{14}
\pubyear{2025}
\copyrightyear{2025}
\externaleditor{Hua~Li} 
\datereceived{9 June 2025}
\daterevised{7 July 2025}
\dateaccepted{15 July 2025}
\datepublished{16 July 2025}
\hreflink{\textls[-30]{https://doi.org/10.3390/en18143769}}
\usepackage[OT1,OT2,T2A,T2B,T2C,T3,T5,T1]{fontenc}
\usepackage[russian,english]{babel}

\Title{Super-Resolution for Renewable Energy Resource Data with Wind from Reanalysis Data and Application to Ukraine}

\TitleCitation{Super-Resolution for Renewable Energy Resource Data with Wind from Reanalysis Data and Application to Ukraine}

\Author{Brandon N.~Benton~*\href{https://orcid.org/0009-0008-9931-2050}{\orcidicon}, Grant~Buster, Pavlo~Pinchuk, Andrew~Glaws \href{https://orcid.org/0000-0002-7268-1883}{\orcidicon}, Ryan N.~King, Galen~Maclaurin \href{https://orcid.org/0000-0001-7491-1614}{\orcidicon} \mbox{and Ilya~Chernyakhovskiy}}

\AuthorNames{Brandon N. Benton, Grant Buster, Pavlo Pinchuk, Andrew Glaws, Ryan N. King, Galen Maclaurin and Ilya Chernyakhovskiy}

\AuthorCitation{Benton, B.N.; Buster, G.; Pinchuk, P.; Glaws, A.; King, R.N.; Maclaurin, G.; Chernyakhovskiy, I.}

\address[1]{National Renewable Energy Laboratory, Golden, CO 80401, USA}

\corres{\hangafter=1 \hangindent=1.0em \hspace{-1em} Correspondence: brandon.benton@nrel.gov}

\abstract{With a potentially increasing share of the electricity grid relying on wind to provide generating capacity and energy, there is an expanding global need for historically accurate, spatiotemporally continuous, high-resolution wind data. Conventional downscaling methods for generating these data based on numerical weather prediction have a high computational burden and require extensive tuning for historical accuracy. In this work, we present a novel deep learning-based spatiotemporal downscaling method using generative adversarial networks (GANs) for generating historically accurate high-resolution wind resource data from the European Centre for Medium-Range Weather Forecasting Reanalysis version 5 data (ERA5). In contrast to previous approaches, which used coarsened high-resolution data as low-resolution training data, we use true low-resolution simulation outputs. We show that by training a GAN model with ERA5 as the low-resolution input and Wind Integration National Dataset Toolkit (WTK) data as the high-resolution target, we achieved results comparable in historical accuracy and spatiotemporal variability to conventional dynamical downscaling. This GAN-based downscaling method additionally reduces computational costs over dynamical downscaling by two orders of magnitude. We applied this approach to downscale 30 km, hourly ERA5 data to 2 km, 5 min wind data for January 2000 through December 2023 at multiple hub heights over Ukraine, Moldova, and part of Romania. With WTK coverage limited to North America from 2007--2013, this is a significant spatiotemporal generalization. The geographic extent centered on Ukraine was motivated by stakeholders and energy-planning needs to rebuild the Ukrainian power grid in a decentralized manner. This 24-year data record is the first member of the super-resolution for renewable energy resource data with wind from the reanalysis data dataset (Sup3rWind).}


\makeatletter
\DeclareRobustCommand*\textsubscript[1]{%
  \@textsubscript{\selectfont#1}}
\def\@textsubscript#1{%
  {\m@th\ensuremath{_{\mbox{\fontsize\sf@size\z@#1}}}}}
\makeatother

\usepackage{sansmath}

\begin{document}
\section{Introduction \label{sect:sec1-energies-3720435}}

With the potential increase in wind energy in the power system, high-resolution spatiotemporal wind resource datasets are becoming increasingly important~\cite{B1-energies-3720435,B2-energies-3720435}. These historically accurate meteorological data are invaluable for ensuring resource adequacy~\cite{B3-energies-3720435}, reliable system operations~\cite{B3-energies-3720435}, well-functioning electricity markets~\cite{B3-energies-3720435}, and more. These applications require wind resource data that capture detailed meteorological processes (i.e., processes occurring at $\leq$3 km and sub-hourly resolutions~\cite{B4-energies-3720435}). Although these data are vital to the success of future investment in wind energy, said data are difficult to produce and rarely available~\cite{B2-energies-3720435}. Purchasing high-resolution time series wind resource data can be costly for large geographic extents covering a long-term historical record, and generating regional or national high-resolution datasets can be expensive in terms of both labor hours and computational costs. Furthermore, the uncertainty of existing high-resolution wind resource data is often not quantified, nor is the data extensively validated against observations~\cite{B1-energies-3720435,B2-energies-3720435}.

The most common approach for generating high-resolution historical meteorological datasets is downscaling global reanalysis data. The downscaling techniques can be roughly separated into two groups: dynamical or statistical downscaling. Dynamical downscaling uses regional climate models or numerical weather prediction models, with lower-resolution meteorological data as lateral boundary conditions, to perform direct numerical simulations of high-resolution fields. Statistical downscaling generates high-resolution fields by applying previously identified statistical relationships between large-scale and small-scale content. Statistical downscaling is computationally efficient but fails to resolve important small-scale features~\cite{B5-energies-3720435}. Dynamical downscaling provides more realistic dynamics, especially over complex terrain, but can be prohibitively expensive to perform over large regions and time periods~\cite{B6-energies-3720435}. Additionally, producing high-fidelity output can require meticulously tailoring dynamical downscaling simulation configurations to the specific application~\cite{B7-energies-3720435,B8-energies-3720435,B9-energies-3720435}.

\subsection{Previous Work \label{sect:sec1dot1-energies-3720435}}

Historically, most spatiotemporally continuous high-resolution wind data have been generated with dynamical downscaling~\cite{B4-energies-3720435,B10-energies-3720435,B11-energies-3720435,B12-energies-3720435} or statistical downscaling~\cite{B13-energies-3720435,B14-energies-3720435,B15-energies-3720435}. As mentioned previously, dynamical downscaling leverages non-linear numerical weather prediction, like the Weather Research and Forecasting System~\cite{B16-energies-3720435}, to generate highly accurate outputs but can require resources that make large-scale data generation infeasible~\cite{B17-energies-3720435}. Dynamical downscaling can also require extensive tuning to select the best physics schemes and constant reinitializations of the simulations to ensure limited drift from the boundary conditions~\cite{B10-energies-3720435,B11-energies-3720435,B18-energies-3720435}. There are numerous statistical downscaling methods such as localized constructed analogs (LOCA)~\cite{B19-energies-3720435}, combined bias correction with spatial disaggregation~\cite{B20-energies-3720435}, and Bartlett--Lewis rectangular pulse models~\cite{B21-energies-3720435}. These methods can be significantly faster than dynamical downscaling but can also fail to simultaneously capture short-time and fine-scale spatial dynamics essential for accurate downstream modeling for energy systems~\cite{B5-energies-3720435}.

The intersection of deep learning and meteorological modeling is an active area of research, with promising developments specifically regarding weather forecasting~\cite{B22-energies-3720435,B23-energies-3720435,B24-energies-3720435,B25-energies-3720435}. Machine learning methods are being adopted by established forecasting centers and will soon play an integral role in operational predictions~\cite{B26-energies-3720435}. However, research in deep learning-based downscaling, also called super-resolution, is less active, especially when it comes to fully gridded spatiotemporal wind speed downscaling. Existing research on machine learning applications to downscaling is mostly focused on pointwise spatial enhancement~\cite{B27-energies-3720435,B28-energies-3720435} with regression methods~\cite{B29-energies-3720435,B30-energies-3720435}, and often for less dynamic fields like precipitation and temperature rather than wind speed~\cite{B28-energies-3720435,B30-energies-3720435,B31-energies-3720435,B32-energies-3720435,B33-energies-3720435}. When wind fields are downscaled, they typically provide a coarse sampling of over a kilometer along the vertical dimension or a single near-surface field~\cite{B27-energies-3720435,B28-energies-3720435,B34-energies-3720435}. However, wind energy modeling applications such as the Renewable Energy Potential model~\cite{B35-energies-3720435} require a finer sampling of near-surface wind fields over typical wind turbine height (s).

Super-resolution leverages deep convolutional networks with various model architectures, such as UNets, CNNs, and GANs~\cite{B36-energies-3720435,B37-energies-3720435,B38-energies-3720435,B39-energies-3720435}. GANs, in particular, have demonstrated superior performance over these standard regression models in generating more realistic spatial structures~\cite{B40-energies-3720435,B41-energies-3720435}. This previous work on super-resolution has relied on the assumption that coarsened or averaged, high-resolution data are a good approximation for low-resolution data~\cite{B37-energies-3720435,B40-energies-3720435,B42-energies-3720435,B43-energies-3720435,B44-energies-3720435}. We previously used GANs trained in this way to downscale wind data over Southeast Asia~\cite{B45-energies-3720435}. While this assumption can lead to excellent results, our approach instead uses low-resolution European Centre for Medium-Range Weather Forecasting Reanalysis v5 (ERA5) input data paired with high-resolution dynamical downscaling outputs as target data. We also include multiple low-resolution variables in training, which are not super-resolved, solely to better inform the enhancement of the high-resolution outputs and improve model generalization. This aligns with the process of dynamical downscaling, which uses multiple variables from low-resolution simulation output as boundary conditions for high-resolution simulations. We show that by training with separate low-resolution and high-resolution simulation data in this way, we achieved performance comparable to dynamical downscaling.

\subsection{Overview \label{sect:sec1dot2-energies-3720435}}

In this work, we present a novel deep learning-based spatiotemporal downscaling approach using generative adversarial networks (GANs). These networks were trained with pairs of low-resolution simulation data from ERA5 and high-resolution simulation data from the Wind Integration National Dataset Toolkit (WTK)~\cite{B10-energies-3720435}. This differs from previous approaches~\cite{B45-energies-3720435}, which use coarsened high-resolution data as the low-resolution training data. With this paired training approach, models learn a transformation closer to dynamical downscaling instead of an un-coarsening operation. Since true low-resolution simulations (ERA5) can differ significantly from coarsened dynamically downscaled data, this approach leads to more accurate and physically realistic outputs. Additionally, when training on coarsened high-resolution data, low-resolution training features must be available in the high-resolution data. By pairing ERA5 and WTK, additional low-resolution training features can be included, which enables models to learn a more robust relationship between the low-resolution climate representation and high-resolution outputs. Fully trained models can then generate accurate high-resolution data from low-resolution input~\cite{B40-energies-3720435} orders of magnitude faster than conventional dynamical downscaling. These models can be deployed for new regions without additional tuning. This deployment is faster and simpler in practice than dynamical downscaling, without many of the logistical difficulties involved, like consistent reinitializations.

We demonstrate the performance of our approach across out-of-sample regions in North America and over Ukraine, Moldova, and parts of Romania. While the primary focus was Ukraine, surrounding areas were included to preserve a rectangular grid. Results from a broad suite of performance measures show excellent fidelity with observations across diverse regions with complex terrain and with underlying physics of dynamically downscaled data. We downscale and make publicly available 24 years (2000--2023) of 30 km, hourly ERA5 to 2 km, 5 min resolution data over this region. With WTK coverage limited to North America from 2007--2013, this is a significant geographic generalization. This 24-year data record is the first member of the super-resolution for renewable energy resource data with wind from the reanalysis data dataset (Sup3rWind). The focus on Ukraine was motivated by stakeholders and energy-planning needs to rebuild the Ukrainian power grid in a decentralized manner after the conflict with Russia. At the end of 2024, the Ukrainian power grid had lost more than 50\% of pre-war capacity~\cite{B46-energies-3720435}, with nearly 90\% of wind power capacity out of operation~\cite{B47-energies-3720435}. However, the high resilience of decentralized generation from wind, strong policy support, and international investment continue to drive more construction~\cite{B48-energies-3720435}.

This paper is organized as follows. In \cref{sect:sec2dot1-energies-3720435,sect:sec2dot2-energies-3720435}, we describe the general problem of downscaling, define the notation used throughout the paper, and discuss the numerous data sources used in this work. In \cref{sect:sec2dot3-energies-3720435,sect:sec2dot4-energies-3720435,sect:sec2dot5-energies-3720435,sect:sec2dot6-energies-3720435}, we cover our GAN model setup, model training, bias correction, and use of the model in inference. In \mbox{\cref{sect:sec3dot1-energies-3720435,sect:sec3dot2-energies-3720435}}, we look at the physical performance of our downscaling results across various performance measures and compare the results against observations across different regions in North America and Ukraine. In \sect{sect:sec4-energies-3720435}, we discuss possible directions for future work. We conclude with final remarks in \sect{sect:sec5-energies-3720435}.

\section{Materials and Methods \label{sect:sec2-energies-3720435}}

\subsection{Problem Statement and Notation \label{sect:sec2dot1-energies-3720435}}

The problem of downscaling low-resolution data is as follows. Given a low-resolution state $x $, a target spatial enhancement $s $ for each spatial dimension ($s^{2} $ overall), and a target temporal enhancement $t $, we want to find a function $G_{s,t} $ that will take $x $, enhance the spatial dimensions by a factor of $s $ and the temporal dimension by a factor of $t $, and give us a spatiotemporally enhanced high-resolution state $x^{\prime} $. Under some simplifying assumptions, we can decompose $G_{s,t}\left( x \right) $ into separate functions for spatial and temporal enhancement, $G_{1,t}\left( {G_{s,1}\left( x \right)} \right) $. We can further decompose these functions into intermediate enhancement functions if the products of intermediate enhancement factors are equal to $s $ or $t $. The terms introduced here, along with other frequently used terms, are summarized in \tabref{tabref:energies-3720435-t001}.    
    \begin{table}[H]
    
    \caption{A summary of terms used in this paper.}
    \label{tabref:energies-3720435-t001}

\begin{adjustwidth}{-\extralength}{0cm}

\setlength{\cellWidtha}{\fulllength/2-2\tabcolsep-0.9in}
\setlength{\cellWidthb}{\fulllength/2-2\tabcolsep+0.9in}
\scalebox{1}[1]{\begin{tabularx}{\fulllength}{>{\centering\arraybackslash}m{\cellWidtha}>{\centering\arraybackslash}m{\cellWidthb}}
\toprule

\textbf{Terms} & \textbf{Meaning}\\
\cmidrule{1-2}

True low-resolution data & Output of a low-resolution simulation. In contrast to artificial low-resolution data obtained through coarsening high-resolution simulation output. The primary example used is ERA5.\\
\cmidrule{1-2}
High-resolution target data $\left( y_{true} \right. $) & Output of a high-resolution dynamical downscaling simulation. In contrast to synthetic high-resolution data obtained through GAN-based downscaling. The primary example used is WTK.\\
\cmidrule{1-2}
$G_{3,1} $ & Generator, trained with ERA5 input and coarsened WTK (10 km, hourly) as target data with modified content loss function, which enhances low-resolution data by spatial factor 3 (first enhancement step).\\
\cmidrule{1-2}
$G_{5,1} $ & Generator, trained with coarsened WTK (10 km, hourly) as input data and subsampled WTK (2 km, hourly) as target data, which enhances low-resolution data by spatial factor 5 (second enhancement step).\\
\cmidrule{1-2}
$G_{1,12} $ & Generator, trained with subsampled WTK (2 km, hourly) as input data and original WTK (2 km, 5 min) as target data, which enhances low-resolution data by temporal factor 12 (third enhancement step).\\
\cmidrule{1-2}
$G_{15,12} $ & Composite generator that performs all three enhancement steps to go from ERA5 (30 km, hourly) to 2 km, 5 min resolution.\\

\bottomrule
\end{tabularx}}

\end{adjustwidth}
    
    \end{table}

\subsection{Data Description \label{sect:sec2dot2-energies-3720435}}

ERA5: We downloaded ERA5~\cite{B49-energies-3720435} for 2007--2013 to train the first enhancement step model. ERA5 is an atmospheric reanalysis dataset that is an optimal combination of observations from various measurement sources and the output of a numerical model using a Bayesian estimation process called data assimilation~\cite{B50-energies-3720435}. ERA5 consists of hourly estimates of several atmospheric variables at a latitude and longitude resolution of 0.25$^\circ$ (\textasciitilde{}30 km at the equator) from the surface of the earth to roughly 100 km altitude from 1979 to the present day.

As our focus is to generate high-resolution wind resource data, we selected variables from ERA5 close to the surface. We also selected variables that would encourage accuracy during extreme events and over different types of complex terrain. Good model generalization also requires learning the relationships between the low-resolution climate representation and the high-resolution outputs. Prior to training, ERA5 data were regridded to match the 15-times spatially coarsened WTK grid, and ERA5 wind components were bias-corrected to the WTK so that the 2007--2013 monthly means and standard deviations matched those of WTK. This ensured that training was not influenced by bias between low- and high-resolution data, and we applied separate bias correction prior to inference. The ERA5 configuration is summarized in \tabref{tabref:energies-3720435-t002}. The complete set of training features used is listed in \tabref{tabref:energies-3720435-t003}.

WTK: WTK is high-resolution (2 km, 5 min) wind data that covers Canada, the United States, and Mexico from 2007 through 2013. We can, in theory, use any ERA-based downscaled data product as high-resolution target data. We selected the National Renewable Energy Laboratory (NREL)’s WTK~\cite{B10-energies-3720435} because of its extensive use by U.S. stakeholders for wind resource and energy production analysis and because WTK has demonstrated good performance across various performance measures. In particular, WTK shows good agreement with observations for diurnal and seasonal correlation coefficients, mean absolute error (MAE), mean wind speeds, and absolute bias~\cite{B51-energies-3720435}. The WTK was produced with Weather Research and Forecasting (WRF) version 3.4.1 using ERA-Interim, the predecessor to ERA5, for initialization and boundary conditions. WTK data include wind speed and wind direction at 10, 40, 80, 100, 120, 160, and 200 m above ground level. The wind speed and wind direction data served as the high-resolution targets for our downscaling framework. Coarsened WTK data are also used as the low-resolution data for $G_{5,1} $ and $G_{1,12} $ (ERA5 data are only needed for input to the first enhancement step). The WTK configuration is summarized in \tabref{tabref:energies-3720435-t002}.

Vortex Wind Data from the International Renewable Energy Agency Global Atlas: We download long-term monthly wind speed means from the International Renewable Energy Agency Global Atlas (data provided by Vortex~\cite{B12-energies-3720435}) over Ukraine and the contiguous United States (CONUS) to use for bias correction prior to inference. Vortex via the International Renewable Energy Agency Global Atlas provides high-resolution wind speed data globally~\cite{B52-energies-3720435} and easily downloadable 20-year climatological monthly means. We bias-corrected ERA5 data over Ukraine by matching the corrected ERA5 monthly means over 2000--2020 with the Vortex monthly means. We bias-corrected ERA5 data over CONUS prior to inference used for validation against observational data. Bias correction is described in \sect{sect:sec2dot5-energies-3720435}.

Meteorological Assimilation Data Ingest System (MADIS): MADIS is a comprehensive collection of meteorological observations covering the entire globe~\cite{B53-energies-3720435}. It is maintained by the National Oceanic and Atmospheric Administration and is primarily used for weather forecasting, research, and various atmospheric studies. MADIS integrates data from various sources, including federal agencies, research institutions, and commercial entities, ensuring broad coverage and diversity of observations. The dataset undergoes quality control procedures to identify and correct errors, ensuring high-quality data for analysis and modeling purposes.

\begin{figure}[H]
      \includegraphics[scale=1]{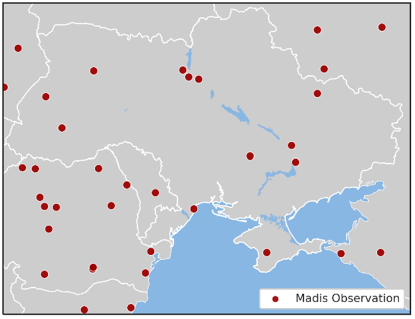}
\caption{Ukraine, Moldova, and Romania downscaling domain. MADIS observation sites are shown in dark red. Wind farm locations are not shown due to security concerns.}
\label{fig:energies-3720435-f001}
\end{figure}

We used the MADIS API to download a full year of surface observations of wind speed and direction for 40 locations within the Ukraine downscaling domain (\fig{fig:energies-3720435-f001}). The observations for each location were mapped onto an hourly temporal grid using a simple average for time steps containing multiple observations. We removed any locations missing observational data for more than half of the time steps. The resulting validation data consists of 8784 hourly observations of wind speed for 2020 at 10 m height for \mbox{37 locations} across the modeling domain.    

Second Wind Forecast Improvement Project (WFIP2): We used WFIP2 observation data to assess model performance over CONUS. WFIP2 is a U.S. Department of Energy and National Oceanic and Atmospheric Administration-funded effort to improve weather prediction forecast skills for turbine-height winds in regions with complex terrain. A core component of WFIP2 was an 18-month field campaign that took place in the U.S. Pacific Northwest between October 2015 and March 2017~\cite{B54-energies-3720435}.

Ukraine Wind Farm Observations: We obtained wind measurement data performed by Deutsche WindGuard Consulting GmbH (Varel, Germany), GEIO-NET Umweltconsulting GmbH (Hannover, Germany), and ENERPARK Inżyniera Wiatrowa Sp. z o. o (Warsaw, Poland) for planned wind farm sites throughout Ukraine. Due to security concerns, we refer to the five wind farm sites as Wind Farm A--E rather than their actual locations. The wind speed measurements for Wind Farm A were conducted using a 100 m high met mast for wind speeds at approximately 100 m and 80 m. Sets of measurements for Wind Farms B and C were performed using a 120 m high met mast, yielding wind speed measurements at approximately 120 m, 100 m, 75 m, and 50 m. The wind speed measurements for Wind Farm D were conducted using a 120 m high met mast for wind speeds at approximately 120 m, 116 m, 100 m, 80 m, and 60 m. Measurements for Wind Farm E were collected using an 82 m high met mast and extrapolated to the turbine hub height of 94 m using wind shear exponents calculated from mast data. This collection of observational wind speeds was used to validate Sup3rWind data across Ukraine (\sect{sect:sec3dot2-energies-3720435}). The Wind Farm observation heights are listed in \tabref{tabref:energies-3720435-t004}.

    \begin{table}[H]
    
    \caption{Summary of ERA5 and wind toolkit configurations.}
    \label{tabref:energies-3720435-t002}
    
{\footnotesize
\begin{adjustwidth}{-\extralength}{0cm}

\setlength{\cellWidtha}{\fulllength/3-2\tabcolsep-0.8in}
\setlength{\cellWidthb}{\fulllength/3-2\tabcolsep+0.4in}
\setlength{\cellWidthc}{\fulllength/3-2\tabcolsep+0.4in}
\scalebox{1}[1]{\begin{tabularx}{\fulllength}{>{\centering\arraybackslash}m{\cellWidtha}>{\centering\arraybackslash}m{\cellWidthb}>{\centering\arraybackslash}m{\cellWidthc}}
\toprule

 & \textbf{ERA5} & \textbf{Wind Toolkit}\\
\cmidrule{1-3}

Output Variables & Numerous meteorological variables at the surface and 137 pressure levels up to around 80 km. Includes wind speed, wind direction, temperature, pressure, relative humidity, heat fluxes, precipitation, cape, etc. For a complete list, see {\cite{B55-energies-3720435}}. & Wind speed, wind direction, air temperature, and pressure at 15 m, 47 m, 80 m, 112 m, 145 m, and 177 m. Interpolated to 10 m, 40 m, 80 m, 100 m, 120 m, 160 m, and 200 m. Surface pressure and relative humidity \mbox{at 2 m.}\\
\cmidrule{1-3}
Resolution & 30 km, hourly. & 2 km, 5 min.\\
\cmidrule{1-3}
Boundary Conditions/Inputs & 4D-Variational Data Assimilation from satellites, surface observations, and other sources. Atmospheric state that best fits model forecast and observations {\cite{B56-energies-3720435}}. Assimilation performed with 12 h windows. & 6-hourly scale-selective grid nudging towards ERA-Interim. GTOPO30 terrain data.\\

\bottomrule
\end{tabularx}}

\end{adjustwidth}}
    
    \end{table}
    \vspace{-10pt}

\subsection{Model Description \label{sect:sec2dot3-energies-3720435}}

\begin{figure}[H]
      \includegraphics[scale=1.3]{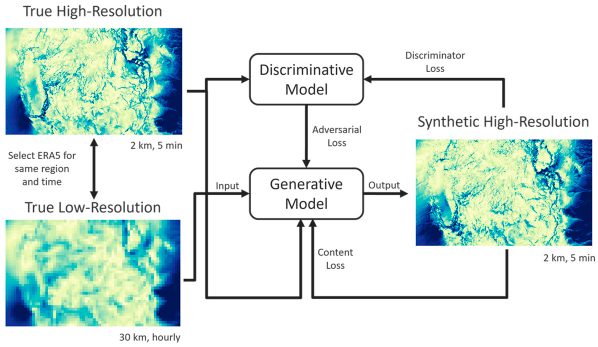}
\caption{GAN training and inference flow. Inference is performed with only the generator.}
\label{fig:energies-3720435-f002}
\end{figure}

For this work, we trained a total of three super-resolution models, described in \mbox{\tabref{tabref:energies-3720435-t003}.} The first step, $G_{3,1} $, performed 3-times spatial enhancement; the second step, $G_{5,1}, $ performed 5-times spatial enhancement; and the third step, $G_{1,12}, $ performed 12-times temporal enhancement. When these steps were applied successively to a low-resolution state $x $, $G_{1,12}\left( {G_{5,1}\left( {G_{3,1}\left( x \right)} \right)} \right) $, they performed a total of 15-times spatial enhancement and 12-times temporal enhancement, $G_{15,12} $. Training and inference flow are diagrammed in \fig{fig:energies-3720435-f002}. These models mostly follow the approach in~\cite{B40-energies-3720435}, with a few important distinctions: (1) we used a modified content loss function to encourage model accuracy across extreme values. This loss, shown in Equation (1), includes mean absolute error terms for the minimums and maximums across both space and time; (2) we incorporated mid-network high-resolution topography injection for a more accurate representation of wind flow over fine-scale complex terrain; and (3) we trained on distinct low-resolution and high-resolution datasets, as opposed to using coarsened high-resolution data as the low-resolution GAN input. The topography injection differs from standard model input in that all standard model inputs are low-resolution. As this low-resolution data goes through the model, it is eventually enhanced by up-sampling layers in the middle of the model network. Right after this up-sampling, high-resolution topography data can be combined with the up-sampled data. This high-resolution topography is elevation above sea level data sourced from GTOPO30~\cite{B57-energies-3720435}.\begin{equation}
\label{eq:FD1-energies-3720435}
\begin{matrix}
{\mathcal{L}\left( {x,y_{true}} \right) =} & {\, mae\left( {y_{true},\ y_{synth}} \right) + mae\left( {\left. {{max}_{t}\left( y \right.}_{true} \right),max_{t}\left( y_{synth} \right)} \right)} \\
 & {+ ~mae\left( {\left. {{min}_{t}\left( y \right.}_{true} \right),{min}_{t}\left( y_{synth} \right)} \right)} \\
 & {+ ~mae\left( {\left. {max_{s}\left( y \right.}_{true} \right),max_{s}\left( y_{synth} \right)} \right)} \\
 & {+ ~mae\left( {\left. {{min}_{s}\left( y \right.}_{true} \right),{min}_{s}\left( y_{synth} \right)} \right)} \\
\end{matrix}
\tag{1}
\end{equation}
    \vspace{-6pt}

The loss function used to encourage accuracy across extreme values, $mae $, is mean absolute value, $y_{true} $ is the true high-resolution data, $y_{synth} $ is the high-resolution model output, $max_{t} $ is the maximum across all time, and $max_{s} $ is the maximum across \mbox{all space.}

An extensive codebase has been developed to implement easily customizable GAN architectures and handle data extraction, batching, and model training to distribute the forward passes of input data through the GAN across multiple nodes. This codebase is released as the super-resolution for renewable resource data (sup3r) package and is installable through the python package index~\cite{B58-energies-3720435}. Sup3r version 0.1.2 was used for \mbox{this work.}

    \begin{table}[H]
    
    \caption{30 km, hourly to 2 km, 5 min model steps.}
    \label{tabref:energies-3720435-t003}
  {\scriptsize  
\begin{adjustwidth}{-\extralength}{0cm}

\setlength{\cellWidtha}{\fulllength/6-2\tabcolsep-0.5in}
\setlength{\cellWidthb}{\fulllength/6-2\tabcolsep-0.2in}
\setlength{\cellWidthc}{\fulllength/6-2\tabcolsep+0.7in}
\setlength{\cellWidthd}{\fulllength/6-2\tabcolsep-0in}
\setlength{\cellWidthe}{\fulllength/6-2\tabcolsep-0in}
\setlength{\cellWidthf}{\fulllength/6-2\tabcolsep-0in}
\scalebox{1}[1]{\begin{tabularx}{\fulllength}{>{\centering\arraybackslash}m{\cellWidtha}>{\centering\arraybackslash}m{\cellWidthb}>{\centering\arraybackslash}m{\cellWidthc}>{\centering\arraybackslash}m{\cellWidthd}>{\centering\arraybackslash}m{\cellWidthe}>{\centering\arraybackslash}m{\cellWidthf}}
\toprule

\textbf{Model Step} & \textbf{Enhancement} & \textbf{Training Features} & \textbf{Input Data Source} & \textbf{Output Target Data Source} & \textbf{Training Time}\\
\cmidrule{1-6}

1. 
        $G_{3,1} $ & Three-times spatial & U/V wind vector components at 10, 100, and 200 m, topography, cape, k index, surface pressure, instantaneous moisture flux, surface temperature, surface latent heat flux, 2 m dewpoint temperature, friction velocity & ERA5 (30 km, hourly) & Coarsened WTK (10 km, hourly) & 240 compute node hours, 2500 epochs\\
\cmidrule{1-6}
2. 
        $G_{5,1} $ & Five-times spatial & U/V wind vector components at 10, 40, 80, 100, 120, 160, and 200 m + topography & Coarsened WTK (10 km, hourly) & Subsampled WTK (2 km, hourly) & 50 compute node hours, 7000 epochs\\
\cmidrule{1-6}
3. 
        $G_{1,12} $ & Twelve-times temporal & U/V wind vector components at 10, 40, 80, 100, 120, 160, and 200 m + topography & Subsampled WTK (2 km, hourly) & Original WTK (2 km, \mbox{5 min}) & 200 compute node hours, 10,000 epochs\\

\bottomrule
\end{tabularx}}
\end{adjustwidth}}

    \end{table}
    \vspace{-10pt}

\subsection{Model Training \label{sect:sec2dot4-energies-3720435}}

The first step generator, $G_{3,1} $, was trained with ERA5 data as low-resolution input and WTK coarsened to 10 km hourly as the high-resolution target for 2007--2009 and 2011--2013. We kept 2010 as a holdout year for validation. The WTK data had a nominal resolution of 2 km, 5 min, so high-resolution targets sampled from these data were coarsened five times spatially and subsampled twelve times temporally for the first model step. Both the second- and third-step models were trained on coarsened WTK data, as in~\cite{B40-energies-3720435}. The input for the second step, $G_{5,1,} $ is 10 km, hourly WTK (five times spatially coarsened and twelve times subsampled in time), and the high-resolution target for $G_{5,1} $ was 2 km, hourly WTK (subsampled 12 times temporally). The input for the third step, $G_{1,12}, $ is WTK subsampled 12 times temporally, and the high-resolution target is the original WTK. These steps are summarized in \tabref{tabref:energies-3720435-t003}.

\begin{figure}[H]
      \includegraphics[scale=1]{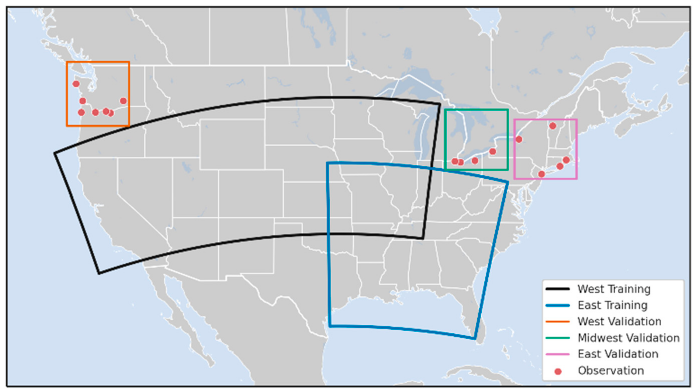}
\caption{GAN training and validation domains. Observation locations outside of training domain shown in red.}
\label{fig:energies-3720435-f003}
\end{figure}

For each model step, training observations were sampled from the domains shown in \fig{fig:energies-3720435-f003}. Training was performed on the Eagle high-performance computing system at NREL using two NVIDIA V100 GPUs (Taiwan Semiconductor Manufacturing Company, Hsinchu Science Park, Taiwan). Each training epoch consisted of 100 batches, with \mbox{64 observations} per batch. Batches were built by randomly sampling spatiotemporal chunks from the six training years and two different training domains. Each spatiotemporal chunk was 15 $\times$ 15 $\times$ 5 low-resolution pixels. For the third step, generator $G_{1,12} $, random sampling along the time dimension was weighted by the time-specific loss. For instance, if the model was performing worst on summer observations during a given training epoch, more observations were selected from the summer for the next epoch. This data-centric training approach ensures that the model performs well over a wide range of season-specific weather conditions.

\subsection{Bias Correction \label{sect:sec2dot5-energies-3720435}}

We performed bias correction on the ERA5 wind speed input data prior to training $G_{3,1} $ and prior to inference. It is well known that ERA5 frequently underestimates wind speeds, especially in complex terrain~\cite{B59-energies-3720435,B60-energies-3720435,B61-energies-3720435}. While the GAN models could be trained on biased data to learn bias correction, we were concerned about this not generalizing well to new geographic regions. Thus, we opted for region-specific bias correction on low-resolution input as a preprocessing step. Prior to training, we computed bias correction factors that shifted the 2007--2013 means and standard deviations of the ERA5 to match those of coarsened WTK data. For each ERA5 grid point ($i,\ j $) and wind speed hub height (10 m, 100 m, or 200 m), monthly ($m $) means ($\upmu$) and standard deviations (\emph{$\sigma$}) were computed for 2007--2013 for both ERA5 and coarsened WTK. ERA5 was then bias-corrected for each grid point, hub height, and month as follows:\begin{equation}
\label{eq:FD2-energies-3720435}
\left. {ERA5}_{ijm}\rightarrow\lbrack{ERA5}_{ijm} - \text{$\upmu$}_{ijm}\rbrack\frac{{\hat{\sigma}}_{ijm}}{\sigma_{ijm}} + {\hat{\text{$\upmu$}}}_{ijm} \right.
\tag{2}
\end{equation}
        where $\text{$\upmu$} $, $\sigma $, ${\hat{{\upmu}}}$, and $\hat{\sigma} $ are the means and standard deviations for ERA5 and the coarsened WTK, respectively.

To perform bias correction prior to inference, we used monthly mean wind speeds provided by Vortex, described in \sect{sect:sec2dot2-energies-3720435}. The global availability of the Vortex data allowed us to use it for both CONUS validation and the Ukraine data production. These means were for 2001--2020 and available only as high as 160 m. Standard deviations were not available. We linearly extrapolated to 200 m, then computed multiplicative correction factors using the means:\begin{equation}
\label{eq:FD3-energies-3720435}
\left. {ERA5}_{ijm}\rightarrow{ERA5}_{ijm}\frac{{\hat{\text{$\upmu$}}}_{ijm}}{\text{$\upmu$}_{ijm}} \right.
\tag{3}
\end{equation}
        where ${\hat{{\upmu}}}$ and $\upmu$ are the 2001--2020 means for Vortex and ERA5, respectively.

\subsection{Inference \label{sect:sec2dot6-energies-3720435}}

We downscaled ERA5 over Ukraine, Moldova, and some of Romania (\fig{fig:energies-3720435-f001}) for 2000--2023, from 30 km hourly to 2 km, 5 min resolution. With models trained only on the CONUS regions shown in \fig{fig:energies-3720435-f003}, this is a significant geographic generalization. Prior to inference, the ERA5 input data were bias-corrected using long-term monthly means from Vortex, described in \sect{sect:sec2dot2-energies-3720435}. Inference is a memory-bound process, so we split the input data into chunks and parallelized the forward pass on these chunks independently. The full low-resolution domain was first chunked across the time dimension, and each chunk, $x, $ passed through $G_{5,1}\left( {G_{3,1}\left( x \right)} \right) $ to perform 15-times spatial enhancement. Chunks were made to overlap in time to enable stitching without seams. Spatially enhanced output was chunked across both space and time, with chunks ($x^{\prime} $) overlapping across all dimensions and then passed through $G_{1,12}\left( x^{\prime} \right) $ to perform the final 12-times temporal enhancement. A year of input for the first two models consisted of 300 chunks. The spatially enhanced input to the final model then consisted of 65,000 spatiotemporal chunks. Forward passes were distributed over 30 compute nodes on the NREL Eagle high-performance computer, and full spatiotemporal enhancement for a year was completed in 40 node hours using \mbox{36 CPUs} per compute node for inference. This is more than 85 times faster than the dynamical downscaling of ERA5 with WRF to the same 2 km, 5 min resolution based on internal testing with WRF on the same hardware. When using GPUs for inference, the speedup can be as much as 500 times.

    \begin{table}[H]
    
    \caption{Wind farm data details.}
    \label{tabref:energies-3720435-t004}

\begin{adjustwidth}{-\extralength}{0cm}

\setlength{\cellWidtha}{\fulllength/3-2\tabcolsep-0in}
\setlength{\cellWidthb}{\fulllength/3-2\tabcolsep-0in}
\setlength{\cellWidthc}{\fulllength/3-2\tabcolsep-0in}
\scalebox{1}[1]{\begin{tabularx}{\fulllength}{>{\centering\arraybackslash}m{\cellWidtha}>{\centering\arraybackslash}m{\cellWidthb}>{\centering\arraybackslash}m{\cellWidthc}}
\toprule

\textbf{Location} & \textbf{Time Period} & \textbf{Heights}\\
\cmidrule{1-3}

Wind Farm A & January 2012--December 2015 & 100 m, 80 m\\
\cmidrule{1-3}
Wind Farm B & September 2019--September 2020 & 120 m, 100 m, 75 m, 50 m\\
\cmidrule{1-3}
Wind Farm C & November 2020--January 2022 & 120 m, 100 m, 75 m, 50 m\\
\cmidrule{1-3}
Wind Farm D & November 2021--September 2023 & 120 m, 116 m, 100 m, 80 m, 60 m\\
\cmidrule{1-3}
Wind Farm E & January 2022--December 2022 & 94 m\\

\bottomrule
\end{tabularx}}
\end{adjustwidth}

    \end{table}
    \vspace{-10pt}

\section{Results \label{sect:sec3-energies-3720435}}

Time and resource limitations prevented extensive hyperparameter search and cross-validation. We trained a few models, with different adversarial weights and selected the one that performed the best on the 2010 WFIP2 observations within the validation regions shown in \fig{fig:energies-3720435-f003}. Model performance was also assessed on 2010 WTK data within these regions. The year 2010 was not included in the training data, and these validation regions were outside of the training domain, so these three regions and time periods enabled spatiotemporal cross-validation. This validation was followed by the generation of a high-resolution 24-year wind data product over Ukraine, Moldova, and eastern Romania. We assessed the performance of these data over Ukraine by comparing them against wind farm and MADIS observational data.

Performance against observations was evaluated with coefficients of determination ($R^{2} $), Pearson correlation coefficients, mean bias error (MBE), MAE, KS-test statistic, diurnal cycle, wind speed variability distribution, bias distribution, and mean relative quantile error (MRQE). $R^{2} $ is defined as the square of the Pearson correlation coefficient, with a value of one indicating that the dependent variable is completely determined by the independent variable and a value of \mbox{zero indicating} the opposite. The KS-test statistic measures the maximum difference between the predicted and empirical CDFs, with a value of \mbox{zero indicating} perfect agreement. The diurnal cycle is the average pattern that occurs over the course of an entire day. The wind speed variability distribution is the probability distribution of the change in wind speed over time. The bias distribution measures the probability of under- or overestimation of wind speed. The MRQE is defined as follows:\begin{equation}
\label{eq:FD4-energies-3720435}
{\text{MRQE} =}\frac{1}{D}{\sum\limits_{i = 1}^{D}\frac{{\hat{Q}}_{i} - Q_{i}}{Q_{i}}}
\tag{4}
\end{equation}
      where ${\hat{Q}}_{i} $ is the $i $-th quantile of the model output, and $Q_{i} $ is the $i $-th quantile of the observation data. We used the MRQE to quantify model performance in resolving extreme events. Negative values indicate underestimation of extremes, and positive values represent overestimation. We evaluated MRQE with 20 logarithmically spaced quantile bins (0.8, 0.999). The MRQE is a particularly important performance measure because accurately capturing long tails is essential for downstream applications of renewable resource data and extreme event estimation. This is also why we compared wind speed variability distributions and KS-test statistics. The wind speed variability distribution is the probability distribution for the wind speed time derivative. The KS-test statistic quantifies the maximum disagreement between cumulative probability distributions of wind speeds.

We estimate the \emph{p}-values for performance measure differences between Sup3rWind and baselines (ERA5 and/or WTK) by bootstrapping distributions for these differences over 1000 samples. For each observation site, we compute the original performance measure difference and the distribution of this performance measure difference by resampling the time series data 1000 times. The proportion of values in this distribution that exceeds the original performance measure difference gives the \emph{p}-value estimate. We additionally compute the \emph{p}-value for time series differences between Sup3rWind and baselines using the Wilcoxon signed-rank test.

\subsection{CONUS Validation \label{sect:sec3dot1-energies-3720435}}

\begin{figure}[H]
      \includegraphics[scale=1]{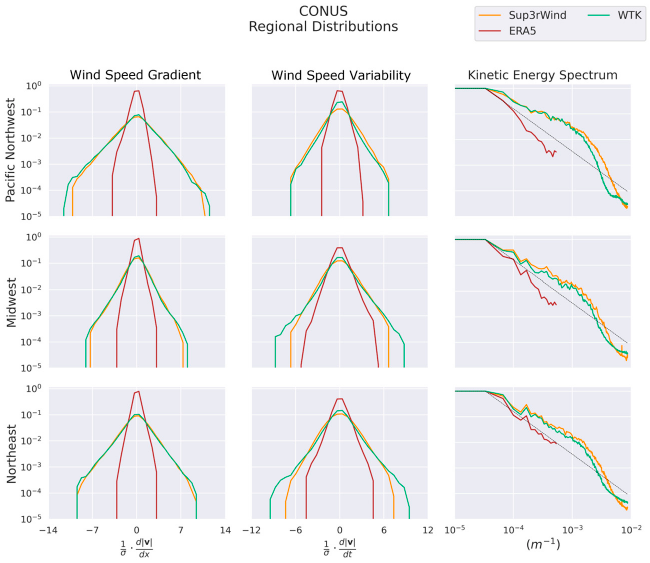}
\caption{Wind speed (100 m AGL) distribution comparisons between ERA5, Sup3rWind, and original WTK across all validation regions. Columns from left to right: probability distribution of longitudinal wind speed gradient, probability distribution of wind speed time derivative, and normalized turbulent kinetic energy spectrum. The dashed line in the kinetic energy plots follows the $k^{- 5/3} $ Kolmogorov scaling law.}
\label{fig:energies-3720435-f004}
\end{figure}

\begin{figure}[H]
          \begin{adjustwidth}{-\extralength}{0cm}
      \centering
      \includegraphics[scale=1]{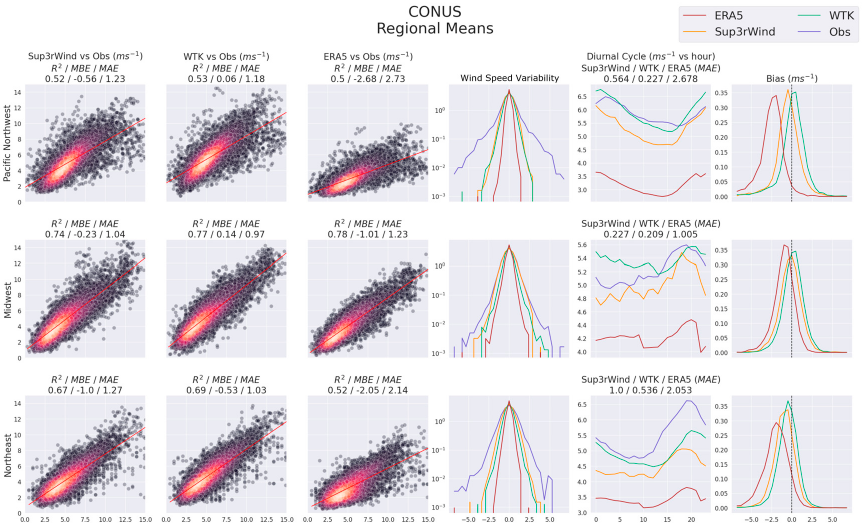}
          \end{adjustwidth}
\caption{Region-wide comparisons against 2010 observations. Columns from left to right: Sup3rWind vs. observation point cloud, WTK vs. observation point cloud, ERA5 vs. observation point cloud, probability distribution of the wind speed variability, diurnal cycle, and bias distribution. Coefficient of determination ($R^{2} $), MAE, and MBE are shown above each scatterplot. The color scheme in the scatter plots is used to show density. The dashed vertical line in the bias distribution plots is positioned at zero bias.}
\label{fig:energies-3720435-f005}
\end{figure}

\fig{fig:energies-3720435-f004} contains calculated statistical and physical quantities for the various validation regions. We see strong agreement between Sup3rWind and WTK; specifically, the long tails of the wind speed gradient and the wind speed variability distribution for the WTK data are well captured by Sup3rWind. Further, the inertial range (i.e., high $k $) region in the turbulent kinetic energy is also recovered by Sup3rWind. In \fig{fig:energies-3720435-f005}, \mbox{Tables \ref{tabref:energies-3720435-t005} and \ref{tabref:energies-3720435-t006}}, we compare Sup3rWind with WFIP2 observations across the three CONUS validation regions. The WFIP2 measurement heights vary by location but are between 20 m and 50 m above ground. Coefficients of determination ($R^{2} $), MAE, and MBE are shown above each scatterplot. For each region, we see excellent agreement between Sup3rWind and WTK and a significant improvement over ERA5. Because we used WTK data for training, we were ultimately limited to the accuracy of this ground truth. There is still room for improvement against observations. We discuss this more in the section Future Research Directions.    
\vspace{-6pt}
    
\vspace{-6pt}
    
    \begin{table}[H]
    
    \caption{Statistics averaged across all CONUS validation regions.}
    \label{tabref:energies-3720435-t005}
    
{\footnotesize
\setlength{\cellWidtha}{\textwidth/4-2\tabcolsep+0.6in}
\setlength{\cellWidthb}{\textwidth/4-2\tabcolsep-0.2in}
\setlength{\cellWidthc}{\textwidth/4-2\tabcolsep-0.2in}
\setlength{\cellWidthd}{\textwidth/4-2\tabcolsep-0.2in}
\scalebox{1}[1]{\begin{tabularx}{\textwidth}{>{\centering\arraybackslash}m{\cellWidtha}>{\centering\arraybackslash}m{\cellWidthb}>{\centering\arraybackslash}m{\cellWidthc}>{\centering\arraybackslash}m{\cellWidthd}}
\toprule

\textbf{Performance Measure} & \textbf{Sup3rWind} & \textbf{WTK} & \textbf{ERA5}\\
\cmidrule{1-4}

MAE & 1.901 m/s & 1.769 m/s & 2.428 m/s\\
\cmidrule{1-4}
MBE & $-$0.434 m/s & 0.079 m/s & $-$1.908 m/s\\
\cmidrule{1-4}
Pearson Correlation Coefficient & 0.721 & 0.741 & 0.692\\
\cmidrule{1-4}
Coefficient of Determination & 0.524 & 0.555 & 0.492\\
\cmidrule{1-4}
Mean Relative Quantile Error & $-$0.075 & $-$0.036 & $-$0.345\\
\cmidrule{1-4}
KS-Test Statistic & 0.115 & 0.109 & 0.292\\

\bottomrule
\end{tabularx}}}

    \end{table}
    \vspace{-8pt}
    
    \begin{table}[H]
    
    \caption{\emph{p}-Values for performance measure differences averaged across all CONUS validation regions.}
    \label{tabref:energies-3720435-t006}
    
{\footnotesize
\setlength{\cellWidtha}{\textwidth/3-2\tabcolsep-0in}
\setlength{\cellWidthb}{\textwidth/3-2\tabcolsep-0in}
\setlength{\cellWidthc}{\textwidth/3-2\tabcolsep-0in}
\scalebox{1}[1]{\begin{tabularx}{\textwidth}{>{\centering\arraybackslash}m{\cellWidtha}>{\centering\arraybackslash}m{\cellWidthb}>{\centering\arraybackslash}m{\cellWidthc}}
\toprule

\textbf{Performance Measure} & \textbf{Sup3rWind vs. WTK} & \textbf{Sup3rWind vs. ERA5}\\
\cmidrule{1-3}

MAE & 0.0379 & 0.0679\\
\cmidrule{1-3}
MBE & 0.0526 & 0.0\\
\cmidrule{1-3}
Pearson Correlation Coefficient & 0.241 & 0.00843\\
\cmidrule{1-3}
Coefficient of Determination & 0.241 & 0.00914\\
\cmidrule{1-3}
Mean Relative Quantile Error & 0.0846 & 0.0\\
\cmidrule{1-3}
Wilcoxon Signed-Rank Test & 0.0373 & 0.0\\

\bottomrule
\end{tabularx}}}

    \end{table}

\subsection{Ukraine, Moldova, and Eastern Romania Performance \label{sect:sec3dot2-energies-3720435}}

We generated 24 total years of wind data over Ukraine, Moldova, and eastern Romania. Using these data for power system modeling requires high resolution, extensive validation, a long-term data record, and physical consistency across a wide range of conditions~\cite{B2-energies-3720435}. We performed extensive validation and demonstration of the accuracy of Sup3rWind with comparisons against data from five wind farm sites and 37 MADIS sites. Some details for the wind farm sites are shown in \tabref{tabref:energies-3720435-t004}. MADIS sites are all 10 m above ground level, and the wind farm data are distributed between 50 m to 130 m above ground level. Performance across MADIS and wind farm sites is comparable to performance across CONUS \mbox{validation regions.}

\subsubsection{Wind Farm Site Comparisons \label{sect:sec3dot2dot1-energies-3720435}}

\begin{figure}[H]
      \includegraphics[scale=1.1]{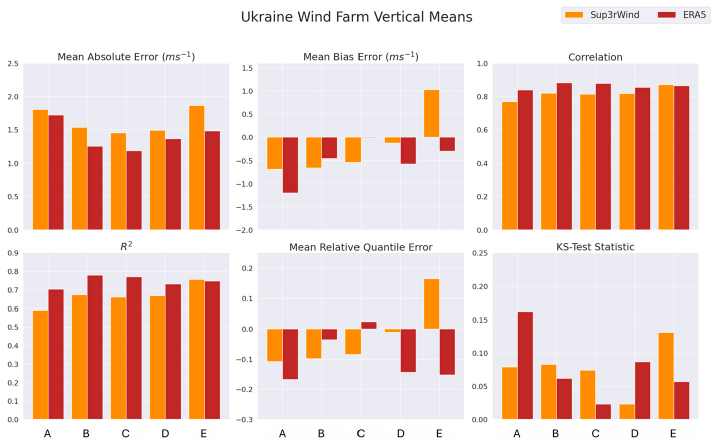}
\caption{Summary of Sup3rWind performance against Ukraine vertically averaged wind farm observations. (\textbf{\boldmath{Top}}), (\textbf{\boldmath{left}}) to (\textbf{\boldmath{right}}): MAE, MBE, and Pearson correlation coefficients. (\textbf{\boldmath{Bottom}}), (\textbf{\boldmath{left}}) to (\textbf{\boldmath{right}}): coefficient of determination, MRQE, and KS-test statistic. A--E labels refer to the wind farms listed in \tabref{tabref:energies-3720435-t004}.}
\label{fig:energies-3720435-f006}
\end{figure}

\begin{figure}[H]
          \begin{adjustwidth}{-\extralength}{0cm}
      \centering
      \includegraphics[scale=1.1]{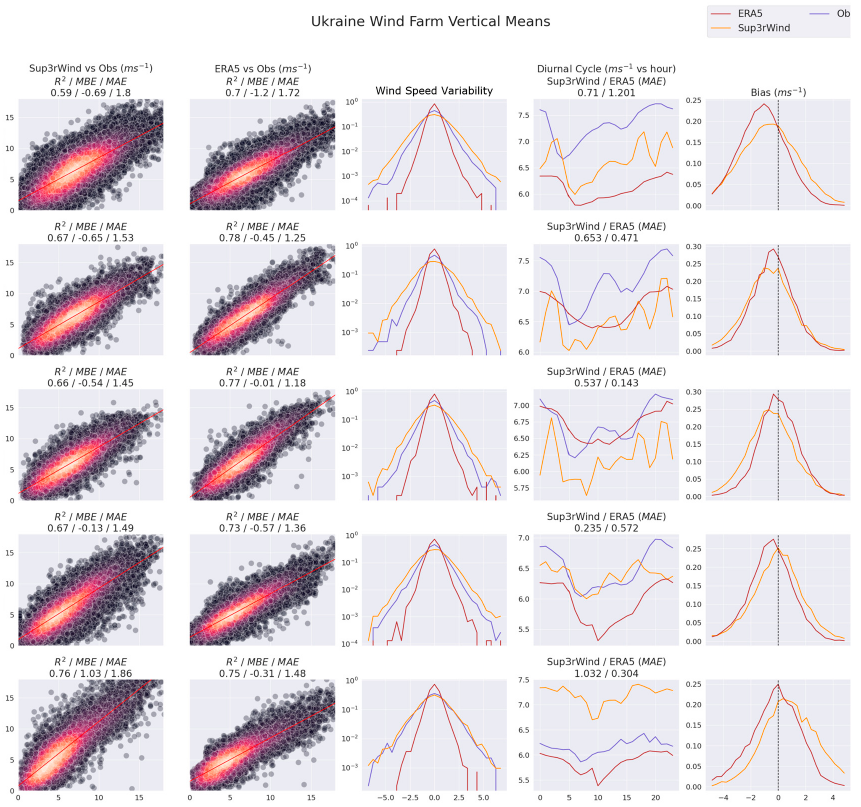}
          \end{adjustwidth}
\caption{Summary of Sup3rWind performance against Ukraine vertically averaged wind farm observations. Columns from left to right: Sup3rWind vs. observation point cloud, ERA5 vs. observation point cloud, probability distribution of the wind speed variability, diurnal cycle, and bias distribution. Wind Farms A--E from top to bottom row. Coefficient of determination ($R^{2} $), MAE, and MBE are shown above each scatterplot. MAE of the diurnal cycle is shown above each diurnal cycle plot. The color scheme in the scatter plots is used to show density. The dashed vertical line in the bias distribution plots is positioned at zero bias.}
\label{fig:energies-3720435-f007}
\end{figure}

In \cref{fig:energies-3720435-f006,fig:energies-3720435-f007}, we show performance against wind farm observations, with each location averaged over all available hub heights. In \fig{fig:energies-3720435-f006}, we see improved MBEs and MRQEs, as well as improvement in KS-test statistics, over ERA5. MBE is within $\pm$1 m per second for each wind farm location. \fig{fig:energies-3720435-f007} shows good agreement with observation for wind speed variability and correlations. We see improvement in MAE for diurnal cycles over ERA5 at some sites, although there is some noise introduced in these cycles, and one site is significantly overestimated. Values of performance measures averaged across all wind farm observations are shown in \tabref{tabref:energies-3720435-t007}. \emph{p}-Values for these performance measures are shown in \tabref{tabref:energies-3720435-t008}.

While CONUS validation showed substantial improvement over ERA5 for Sup3rWind, we do not see the same relative performance across Ukraine. Statistics for Sup3rWind in Ukraine fall in a similar range as for CONUS, while ERA5 performs significantly better. Sup3rWind provides the most improvement here on spatiotemporal variability, relative quantile errors, and KS-test statistics. The increased performance of ERA5 is likely due to the less complex terrain. In the CONUS validation, we saw the best performance of ERA5 in the Midwest, the flattest region. We also saw the best correlations between Sup3rWind and Wind Farm E, the site closest to the Carpathian Mountains.

\vspace{-6pt}
    
\vspace{-6pt}
    
    \begin{table}[H]
    
    \caption{Statistics averaged across all wind farm observations.}
    \label{tabref:energies-3720435-t007}

\setlength{\cellWidtha}{\textwidth/3-2\tabcolsep+0.4in}
\setlength{\cellWidthb}{\textwidth/3-2\tabcolsep-0.2in}
\setlength{\cellWidthc}{\textwidth/3-2\tabcolsep-0.2in}
\scalebox{1}[1]{\begin{tabularx}{\textwidth}{>{\centering\arraybackslash}m{\cellWidtha}>{\centering\arraybackslash}m{\cellWidthb}>{\centering\arraybackslash}m{\cellWidthc}}
\toprule

\textbf{Performance Measure} & \textbf{Sup3rWind} & \textbf{ERA5}\\
\cmidrule{1-3}

MAE & 1.7186 m/s & 1.6202 m/s\\
\cmidrule{1-3}
MBE & $-$0.4879 m/s & $-$0.7407 m/s\\
\cmidrule{1-3}
Pearson Correlation Coefficient & 0.7598 & 0.8016\\
\cmidrule{1-3}
Coefficient of Determination & 0.5772 & 0.6426\\
\cmidrule{1-3}
MRQE & $-$0.105 & $-$0.1321\\
\cmidrule{1-3}
KS-Test Statistic & 0.0671 & 0.1124\\

\bottomrule
\end{tabularx}}

    \end{table}
    \vspace{-6pt}
    
    \begin{table}[H]
    
    \caption{\emph{p}-Values for performance measure differences averaged across all wind farm observations.}
    \label{tabref:energies-3720435-t008}

\setlength{\cellWidtha}{\textwidth/2-2\tabcolsep-0in}
\setlength{\cellWidthb}{\textwidth/2-2\tabcolsep-0in}
\scalebox{1}[1]{\begin{tabularx}{\textwidth}{>{\centering\arraybackslash}m{\cellWidtha}>{\centering\arraybackslash}m{\cellWidthb}}
\toprule

\textbf{Performance Measure} & \textbf{Sup3rWind vs. ERA5}\\
\cmidrule{1-2}

MAE & 0.0304\\
\cmidrule{1-2}
MBE & 0.041\\
\cmidrule{1-2}
Pearson Correlation Coefficient & 0.0378\\
\cmidrule{1-2}
Coefficient of Determination & 0.0378\\
\cmidrule{1-2}
MRQE & 0.0\\
\cmidrule{1-2}
Wilcoxon Signed-Rank Test & 0.00158\\

\bottomrule
\end{tabularx}}

    \end{table}

\subsubsection{MADIS Site Comparisons \label{sect:sec3dot2dot2-energies-3720435}}

\begin{figure}[H]
          \begin{adjustwidth}{-\extralength}{0cm}
      \centering
      \includegraphics[scale=1]{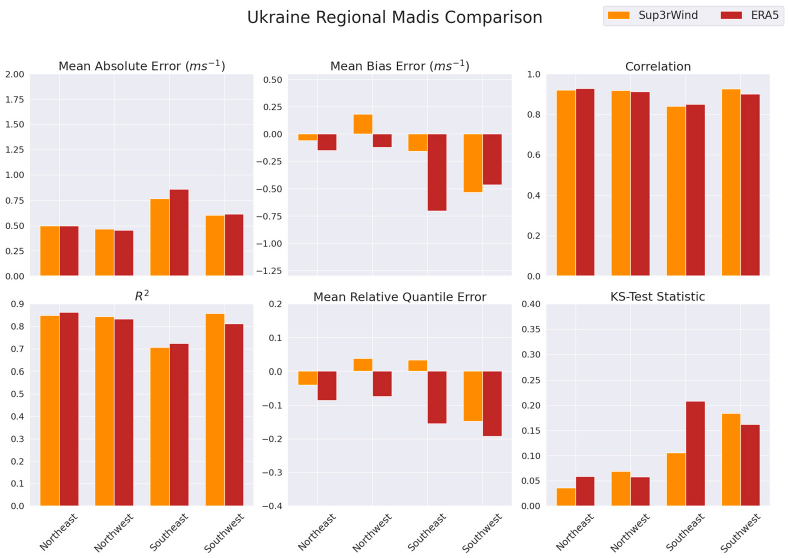}
          \end{adjustwidth}
\caption{Summary of performance against Ukraine MADIS observations. (\textbf{\boldmath{Top}}), (\textbf{\boldmath{left}}) to (\textbf{\boldmath{right}}): MAE, MBE, and Pearson correlation coefficients. (\textbf{\boldmath{Bottom}}), (\textbf{\boldmath{left}}) to (\textbf{\boldmath{right}}): coefficient of determination, MRQE, and KS-test statistic.}
\label{fig:energies-3720435-f008}
\end{figure}

\begin{figure}[H]
          \begin{adjustwidth}{-\extralength}{0cm}
      \centering
      \includegraphics[scale=1]{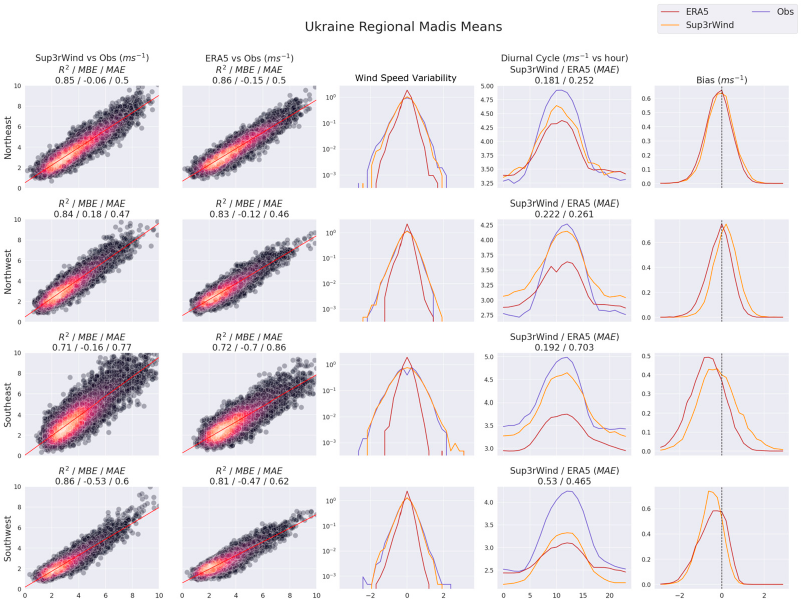}
          \end{adjustwidth}
\caption{Summary of Sup3rWind performance against Ukraine MADIS observations. Columns from left to right: Sup3rWind vs. observation point cloud, ERA5 vs. observation point cloud, probability distribution of the wind speed variability, diurnal cycle, and bias distribution. Coefficient of determination ($R^{2} $), MAE, and MBE are shown above each scatterplot. MAE of the diurnal cycle is shown above each diurnal cycle plot. The color scheme in the scatter plots is used to show density. The dashed vertical line in the bias distribution plots is positioned at zero bias.}
\label{fig:energies-3720435-f009}
\end{figure}

We additionally look at the performance of Sup3rWind across multiple MADIS sites. MADIS measurements are near-surface, approximately 10 m above ground level. It is important to note that near-surface performance can differ significantly from performance at typical wind turbine height. To summarize performance across many MADIS sites, we computed statistics on regional averages. Each of the four quadrants of the spatial domain was used to compute northeast, southeast, southwest, and northwest regional averages. Performance relative to ERA5 for these regions is shown in \cref{fig:energies-3720435-f008,fig:energies-3720435-f009}. In \fig{fig:energies-3720435-f008}, we see excellent agreement with observations, with high correlations and MBE within \mbox{$\pm$1 m} per second for all regions. We also see better performance in capturing extreme values, as measured with MRQE. Values averaged over all MADIS sites are shown in \tabref{tabref:energies-3720435-t009}. Statistics averaged across all MADIS sites. The associated \emph{p}-values are shown in \tabref{tabref:energies-3720435-t010}. In \fig{fig:energies-3720435-f009}, we see improved wind speed variability distributions and diurnal cycles. We again see good performance for ERA5 across the region. The most favorable comparison between Sup3rWind and ERA5 for correlations is seen in the southwest, where the terrain is most complex.    
\vspace{-6pt}
    
\vspace{-6pt}
    
    \begin{table}[H]
    
    \caption{Statistics averaged across all MADIS sites.}
    \label{tabref:energies-3720435-t009}

\setlength{\cellWidtha}{\textwidth/3-2\tabcolsep+0.4in}
\setlength{\cellWidthb}{\textwidth/3-2\tabcolsep-0.2in}
\setlength{\cellWidthc}{\textwidth/3-2\tabcolsep-0.2in}
\scalebox{1}[1]{\begin{tabularx}{\textwidth}{>{\centering\arraybackslash}m{\cellWidtha}>{\centering\arraybackslash}m{\cellWidthb}>{\centering\arraybackslash}m{\cellWidthc}}
\toprule

\textbf{Performance Measure} & \textbf{Sup3rWind} & \textbf{ERA5}\\
\cmidrule{1-3}

MAE & 0.4209 m/s & 0.4743 m/s\\
\cmidrule{1-3}
MBE & $-$0.1453 m/s & $-$0.2389 m/s\\
\cmidrule{1-3}
Pearson Correlation Coefficient & 0.9088 & 0.8999\\
\cmidrule{1-3}
Coefficient of Determination & 0.8259 & 0.8098\\
\cmidrule{1-3}
MRQE & $-$0.0543 & $-$0.1287\\
\cmidrule{1-3}
KS-Test Statistic & 0.0598 & 0.1011\\

\bottomrule
\end{tabularx}}

    \end{table}
    \vspace{-6pt}
    
    \begin{table}[H]
    
    \caption{\emph{p}-Values for performance measure differences averaged across all MADIS sites.}
    \label{tabref:energies-3720435-t010}

\setlength{\cellWidtha}{\textwidth/2-2\tabcolsep-0in}
\setlength{\cellWidthb}{\textwidth/2-2\tabcolsep-0in}
\scalebox{1}[1]{\begin{tabularx}{\textwidth}{>{\centering\arraybackslash}m{\cellWidtha}>{\centering\arraybackslash}m{\cellWidthb}}
\toprule

\textbf{Performance Measure} & \textbf{Sup3rWind vs. ERA5}\\
\cmidrule{1-2}

MAE & 0.00651\\
\cmidrule{1-2}
MBE & 0.0212\\
\cmidrule{1-2}
Pearson Correlation Coefficient & 0.0167\\
\cmidrule{1-2}
Coefficient of Determination & 0.0165\\
\cmidrule{1-2}
MRQE & 0.0298\\
\cmidrule{1-2}
Wilcoxon Signed-Rank Test & 0.00211\\

\bottomrule
\end{tabularx}}

    \end{table}

\section{Discussion \label{sect:sec4-energies-3720435}}

The results shown in this paper support the use of the wind data created by GAN-based downscaling. Downscaling ERA5 data with GANs is shown to produce physically realistic wind across space and time (\fig{fig:energies-3720435-f004}) and historically accurate profiles when compared to ground measurements (\cref{fig:energies-3720435-f005,fig:energies-3720435-f007,fig:energies-3720435-f009}) in nearly all out-of-sample validation conditions. This approach is shown to generalize well to different geographic regions, with training data selected only from CONUS and inference performed over Eastern Europe. Conditioning model output on high-resolution terrain data, a broad set of low-resolution features, and region-specific bias correction should enable the model to generalize to arbitrary regions. While a year-long 2 km, 5 min WRF simulation for CONUS is estimated to cost \mbox{50,000 compute} node hours on the NREL high-performance computing hardware, our GAN framework can create a year of equivalent high-resolution data in 585 compute node hours using CPUs for inference. This shows a more than 85-times speedup for GAN-based downscaling over dynamical downscaling with WRF. The speedup can be as much as \mbox{500 times} when using GPUs for inference.

We see good agreement between Sup3rWind, WTK, and observations across a broad suite of performance measures. Through the probability distributions for the temporal derivative and spatial gradient of wind speed and the turbulent kinetic energy spectrum, we see that Sup3rWind achieves excellent fidelity for the underlying physics of the high-resolution target data. Through site-specific coefficients of determination, absolute errors, and bias errors, we see high fidelity between Sup3rWind and observations across diverse regions with complex terrain.

Our efforts culminated in the production of a 24-year wind data record, with 2 km, \mbox{5 min} spatiotemporal resolution, over Ukraine, Moldova, and eastern Romania. These data were extensively validated using observational data from over 40 different locations, spanning over 9 years, covering heights from 10 m to 120 m above the ground. The performance for Sup3rWind over Ukraine was comparable to that over the CONUS validation regions, showing low mean errors and high correlations. Sup3rWind agreed well with ERA5 while significantly improving the representation of wind speed variability and accuracy of extremes. Diurnal cycles were mostly improved over ERA5, while some noise was introduced in these cycles at wind farm locations. MBE also improved on average, although there is room for improvement on a site-wise basis. All data, models, and software produced through this work are publicly released at no cost, described more in Data \mbox{Availability Statement.}

\subsection*{Future Research Directions }

This work poses a variety of additional research directions to pursue. While the model performance shown here is impressive, especially considering the limited training data and training time, we would like to improve accuracy even further. In the future, we would like to conduct a thorough architecture optimization to reduce network complexity and further speed up inference. Additionally, the models presented in this work were trained using only two GPUs and 6 years of training data. This is extremely limited by industry standards, where weather forecasting models are frequently trained on 30+ years of data and on 100+ GPUs~\cite{B7-energies-3720435,B40-energies-3720435}. Increasing the amount of training data and computational resources could further improve accuracy. Within the confines of the established framework, we are definitionally limited to the accuracy of the high-resolution target dataset. To combat this, we would like to perform a more extensive feature importance analysis on the broad set of ERA5 variables available for training and to explore physics-based loss terms derived from the Navier--Stokes equations. We can leverage some previous work on ERA5 feature importance~\cite{B62-energies-3720435}. Another exciting path for future work would focus on incorporating available observational data as part of either training or post-training data assimilation.

\section{Conclusions \label{sect:sec5-energies-3720435}}

In this work, we have shown that by training a GAN model using ERA5 input data and WTK target data, we achieved results comparable in historical accuracy and spatiotemporal variability to conventional dynamical downscaling. Additionally, we extended the spatial enhancement GAN framework described in~\cite{B40-energies-3720435} to include temporal enhancement, incorporate a modified content loss function to encourage the accuracy of extreme values, and include a mid-network high-resolution topography injection that improved the high-resolution resource assessment in complex terrain. We demonstrated the use and performance of this method through comparisons with high-resolution target data and observational data for CONUS regions in the Pacific Northwest, Midwest, and Northeast. We downscaled ERA5 with this approach to produce a 24-year, high-resolution, high-accuracy, extensively validated wind dataset over Ukraine, Moldova, and eastern Romania. The ERA5 data were enhanced by 15 times along each spatial dimension and 12 times along the temporal dimension, going from 30 km hourly to 2 km, 5 min resolution. These data are comparable to state-of-the-art wind resource datasets developed with physics-based models and are publicly available through multiple easy-access options. We saw strong fidelity across performance measures and observation comparisons while reducing computation expense by two orders of magnitude. Python code for feature engineering, data handling, model training, and inference is also publicly available~\cite{B58-energies-3720435}.

\vspace{6pt}
\authorcontributions{B.N.B. developed software, developed methods, trained models, produced data, and wrote the paper. G.B. developed software, advised on methods, and wrote the paper. P.P. developed software and wrote the paper. A.G. and R.N.K. advised on methods and wrote the paper. G.M. and I.C. advised on and wrote the paper. All authors have read and agreed to the published version of the manuscript.}
\funding{This work was authored in part by the National Renewable Energy Laboratory (NREL), operated by Alliance for Sustainable Energy, LLC, for the U.S. Department of Energy (DOE) under Contract No. DE-AC36-08GO28308. Funding provided by the United States Agency for International Development (USAID) under Contract No. IAG-17-2050. The views expressed in this report do not necessarily represent the views of the DOE or the U.S. Government, or any agency thereof, including USAID. The publisher, by accepting the article for publication, acknowledges that the U.S. Government retains a nonexclusive, paid-up, irrevocable, worldwide license to publish or reproduce the published form of this work or allow others to do so for U.S. Government purposes.}
\dataavailability{The software developed for feature engineering, data handling, training, and inference is available on GitHub at \url{https://github.com/NREL/sup3r} (accessed on 14 July 2025). Sup3r version 0.1.2 was specifically used for this work. The full environment yaml file and the configuration files used to run inference are available at \url{https://github.com/NREL/sup3r/tree/main/examples/sup3rwind} (accessed on 14 July 2025). Training data for this work was obtained through the NREL WIND Toolkit, which is available for download from \url{https://www.nrel.gov/grid/wind-toolkit.html} (accessed on 14 October 2022), and ERA5, which is available from \url{https://www.ecmwf.int/en/forecasts/dataset/ecmwf-reanalysis-v5} (accessed on 10 June 2024). The sup3r software also provides utilities for downloading ERA5 data and performing pre-processing. The final data over Ukraine, Moldova, and Romania are easily accessible through NREL’s Renewable Energy Data Explorer (\url{www.re-explorer.org} (accessed on 14 July 2025)). Additionally, NREL provides several API options where users can download the data with Python or other programming languages (more information can be found at \url{https://developer.nrel.gov/docs/wind/wind-toolkit/sup3rwind-ukraine-download} (accessed on 14 July 2025)). The full dataset is available for download directly via the Open Energy Data Initiative on Amazon Web Services Public Datasets at Directly via OEDI on AWS Public Datasets: nrel-pds-wtk/sup3rwind/ukraine/v1.0.0/5 min and \mbox{nrel-pds-wtk/sup3rwind/ukraine/v1.0.0/60 min.}}
\acknowledgments{The authors would like to thank Caroline Draxl, Evan Rosenlieb, Guilherme Pimenta Castelao, and Jaemo Yang for their thoughtful reviews. The authors would also like to thank Reid Olson and Nicole Taverna for making the super-resolution for renewable energy resource data with wind from reanalysis data (Sup3rWind) and models available via the Open Energy \mbox{Data Initiative.}

}
\conflictsofinterest{The authors declare no conflicts of interest.}
\begin{adjustwidth}{-\extralength}{0cm}

\reftitle{References}

\end{adjustwidth}
\begin{adjustwidth}{-\extralength}{0cm}
\PublishersNote{}
\end{adjustwidth}


\begin{thebibliography}{999}
\bibitem{B1-energies-3720435}
Holttinen, H.; Kiviluoma, J.; Levy, T.; Jun, L.; Eriksen, P.B.; Orths, A.; Cutululis, N.; Silva, V.; Neau, E.; Dobschinski, J.;~et~al. \emph{Design and Operation of Power Systems with Large Amounts of Wind Power: Final Summary Report}; IEA WIND Task 25, Phase four 2015--2017, in VTT Technology; VTT Technical Research Centre of Finland: Espoo, Finland, 2019. [\href{https://doi.org/10.32040/2242-122X.2019.T350}{CrossRef}]

\bibitem{B2-energies-3720435}
Sharp, J.; Milligan, M.; Bloomfield, H.C. Weather Dataset Needs for Planning and Analyzing Modern Power Systems. October 2023. Available online: \href{https://www.esig.energy/wp-content/uploads/2023/10/ESIG-Weather-Datasets-full-report-2023b.pdf}{\nolinkurl{https://www.esig.energy/wp-content/uploads/2023/10/ESIG-Weather-Datasets-full-report-2023b.pdf}} (accessed on 14 July 2025).

\bibitem{B3-energies-3720435}
Dong, Z.; Wong, K.P.; Meng, K.; Luo, F.; Yao, F.; Zhao, J. Wind power impact on system operations and planning. In Proceedings of the IEEE PES General Meeting, Minneapolis, MN, USA, 25--29 July 2010; IEEE: New York, NY, USA, 2010; pp. 1--5. Available online: \href{https://ieeexplore.ieee.org/abstract/document/5590222/}{\nolinkurl{https://ieeexplore.ieee.org/abstract/document/5590222/}} (accessed on 27 April 2024).

\bibitem{B4-energies-3720435}
Clifton, A.; Hodge, B.; Draxl, C.; Badger, J.; Habte, A. Wind and solar resource data sets. \emph{WIREs Energy Environ.} \textbf{\boldmath{2018}}, \emph{7}, e276. [\href{https://doi.org/10.1002/wene.276}{CrossRef}]

\bibitem{B5-energies-3720435}
Murphy, J. An Evaluation of Statistical and Dynamical Techniques for Downscaling Local Climate. \emph{J. Clim.} \textbf{\boldmath{1999}}, \emph{12}, 2256--2284. [\href{https://doi.org/10.1175/1520-0442(1999)012\%3C2256:AEOSAD\%3E2.0.CO;2}{CrossRef}]

\bibitem{B6-energies-3720435}
Martinez-García, F.P.; Contreras-de-Villar, A.; Muñoz-Perez, J.J. Review of Wind Models at a Local Scale: Advantages and Disadvantages. \emph{J. Mar. Sci. Eng.} \textbf{\boldmath{2021}}, \emph{9}, 318. [\href{https://doi.org/10.3390/jmse9030318}{CrossRef}]

\bibitem{B7-energies-3720435}
Benton, B.N.; Alessi, M.J.; Herrera, D.A.; Li, X.; Carrillo, C.M.; Ault, T.R. Minor impacts of major volcanic eruptions on hurricanes in dynamically-downscaled last millennium simulations. \emph{Clim. Dyn.} \textbf{\boldmath{2022}}, \emph{59}, 1597--1615. [\href{https://doi.org/10.1007/s00382-021-06057-4}{CrossRef}]

\bibitem{B8-energies-3720435}
Knutson, T.R.; Sirutis, J.J.; Bender, M.A.; Tuleya, R.E. Dynamical Downscaling Projections of Late 21st Century US Landfalling Hurricane Activity. \emph{Clim. Change} {{2021}},  \emph{in press}.

\bibitem{B9-energies-3720435}
Rockel, B.; Castro, C.L.; Pielke Sr, R.A.; von Storch, H.; Leoncini, G. Dynamical downscaling: Assessment of model system dependent retained and added variability for two different regional climate models. \emph{J. Geophys. Res. Atmos.} \textbf{\boldmath{2008}}, \emph{113}, D21107. [\href{https://doi.org/10.1029/2007JD009461}{CrossRef}]

\bibitem{B10-energies-3720435}
Draxl, C.; Clifton, A.; Hodge, B.M.; McCaa, J. The Wind Integration National Dataset (WIND) Toolkit. \emph{Appl. Energy} \textbf{\boldmath{2015}}, \emph{151}, 355--366. [\href{https://doi.org/10.1016/j.apenergy.2015.03.121}{CrossRef}]

\bibitem{B11-energies-3720435}
Draxl, C.; Wang, J.; Sheridan, L.; Jung, C.; Bodini, N.; Buckhold, S.; Aghili, C.; Peco, K.; Kotamarthi, R.; Kumler, A.;~et~al. \emph{WTK-LED: The WIND Toolkit Long-Term Ensemble Dataset}; National Renewable Energy Laboratory: Golden, CO, USA, 2024. [\href{https://doi.org/10.2172/2473210}{CrossRef}]

\bibitem{B12-energies-3720435}
VORTEX FdC, S.L. Vortex ERA5 Downscaling: Validation Results. Available online: \href{https://www.vortexfdc.com/assets/docs/validation_ERA5.pdf}{\nolinkurl{https://www.vortexfdc.com/assets/docs/validation\_ERA5.pdf}} (accessed on 6 April 2024).

\bibitem{B13-energies-3720435}
Winstral, A.; Jonas, T.; Helbig, N. Statistical Downscaling of Gridded Wind Speed Data Using Local Topography. \emph{J. Hydrometeorol.} \textbf{\boldmath{2017}}, \emph{18}, 335--348. [\href{https://doi.org/10.1175/JHM-D-16-0054.1}{CrossRef}]

\bibitem{B14-energies-3720435}
González-Aparicio, I.; Monforti, F.; Volker, P.; Zucker, A.; Careri, F.; Huld, T.; Badger, J. Simulating European wind power generation applying statistical downscaling to reanalysis data. \emph{Appl. Energy} \textbf{\boldmath{2017}}, \emph{199}, 155--168. [\href{https://doi.org/10.1016/j.apenergy.2017.04.066}{CrossRef}]

\bibitem{B15-energies-3720435}
Salameh, T.; Drobinski, P.; Vrac, M.; Naveau, P. Statistical downscaling of near-surface wind over complex terrain in southern France. \emph{Meteorol. Atmos. Phys.} \textbf{\boldmath{2009}}, \emph{103}, 253--265. [\href{https://doi.org/10.1007/s00703-008-0330-7}{CrossRef}]

\bibitem{B16-energies-3720435}
Onwukwe, C.; Jackson, P.L. Meteorological Downscaling with WRF Model, Version 4.0, and Comparative Evaluation of Planetary Boundary Layer Schemes over a Complex Coastal Airshed. \emph{J. Appl. Meteorol. Climatol.} \textbf{\boldmath{2020}}, \emph{59}, 1295--1319. [\href{https://doi.org/10.1175/JAMC-D-19-0212.1}{CrossRef}]

\bibitem{B17-energies-3720435}
Zhou, E.; Mai, T. \emph{Electrification Futures Study: Operational Analysis of U.S. Power Systems with Increased Electrification and Demand-Side Flexibility}; National Renewable Energy Laboratory (NREL): Golden, CO, USA, 2021. [\href{https://doi.org/10.2172/1785329}{CrossRef}]

\bibitem{B18-energies-3720435}
Michalakes, J.; Hacker, J.; Loft, R.; McCracken, M.O.; Snavely, A.; Wright, N.J.; Spelce, T.; Gorda, B.; Walkup, R. WRF nature run. \emph{J. Phys. Conf. Ser.} \textbf{\boldmath{2008}}, \emph{125}, 012022. [\href{https://doi.org/10.1088/1742-6596/125/1/012022}{CrossRef}]

\bibitem{B19-energies-3720435}
Pierce, D.W.; Cayan, D.R.; Thrasher, B.L. Statistical Downscaling Using Localized Constructed Analogs (LOCA). \emph{J. Hydrometeorol.} \textbf{\boldmath{2014}}, \emph{15}, 2558--2585. [\href{https://doi.org/10.1175/JHM-D-14-0082.1}{CrossRef}]

\bibitem{B20-energies-3720435}
Wood, A.W.; Leung, L.R.; Sridhar, V.; Lettenmaier, D.P. Hydrologic Implications of Dynamical and Statistical Approaches to Downscaling Climate Model Outputs. \emph{Clim. Change} \textbf{\boldmath{2004}}, \emph{62}, 189--216. [\href{https://doi.org/10.1023/B:CLIM.0000013685.99609.9e}{CrossRef}]

\bibitem{B21-energies-3720435}
Kaczmarska, J.; Isham, V.; Onof, C. Point process models for fine-resolution rainfall. \emph{Hydrol. Sci. J.} \textbf{\boldmath{2014}}, \emph{59}, 1972--1991. [\href{https://doi.org/10.1080/02626667.2014.925558}{CrossRef}]

\bibitem{B22-energies-3720435}
Bi, K.; Xie, L.; Zhang, H.; Chen, X.; Gu, X.; Tian, Q. Pangu-Weather: A 3D High-Resolution Model for Fast and Accurate Global Weather Forecast. \emph{arXiv} \textbf{\boldmath{2022}}, arXiv:2211.02556. [\href{https://doi.org/10.48550/arXiv.2211.02556}{CrossRef}]

\bibitem{B23-energies-3720435}
Lam, R.; Sanchez-Gonzalez, A.; Willson, M.; Wirnsberger, P.; Fortunato, M.; Alet, F.; Ravuri, S.; Ewalds, T.; Eaton-Rosen, Z.; Hu, W.;~et~al. Learning skillful medium-range global weather forecasting. \emph{Science} \textbf{\boldmath{2023}}, \emph{382}, 1416--1421. [\href{https://doi.org/10.1126/science.adi2336}{CrossRef}] [\href{https://www.ncbi.nlm.nih.gov/pubmed/37962497}{PubMed}]

\bibitem{B24-energies-3720435}
Nguyen, T.; Brandstetter, J.; Kapoor, A.; Gupta, J.K.; Grover, A. ClimaX: A foundation model for weather and climate. \emph{arXiv} \textbf{\boldmath{2023}}. [\href{https://doi.org/10.48550/arXiv.2301.10343}{CrossRef}]

\bibitem{B25-energies-3720435}
Pathak, J.; Subramanian, S.; Harrington, P.; Raja, S.; Chattopadhyay, A.; Mardani, M.; Kurth, T.; Hall, D.; Li, Z.; Azizzadenesheli, K.;~et~al. FourCastNet: A Global Data-driven High-resolution Weather Model using Adaptive Fourier Neural Operators. \emph{arXiv} \textbf{\boldmath{2022}}, arXiv:2202.11214. [\href{https://doi.org/10.48550/arXiv.2202.11214}{CrossRef}]

\bibitem{B26-energies-3720435}
Morrissey, M. ECMWF Unveils Alpha Version of New ML Model, ECMWF. Available online: \href{https://www.ecmwf.int/en/about/media-centre/aifs-blog/2023/ECMWF-unveils-alpha-version-of-new-ML-model}{\nolinkurl{https://www.ecmwf.int/en/about/media-centre/aifs-blog/2023/ECMWF-unveils-alpha-version-of-new-ML-model}} (accessed on 27 March 2024).

\bibitem{B27-energies-3720435}
Gerges, F.; Boufadel, M.C.; Bou-Zeid, E.; Nassif, H.; Wang, J.T.L. Downscaling daily wind speed with Bayesian deep learning for climate monitoring. \emph{Int. J. Data Sci. Anal.} \textbf{\boldmath{2023}}, \emph{17}, 411--424. [\href{https://doi.org/10.1007/s41060-023-00397-6}{CrossRef}]

\bibitem{B28-energies-3720435}
Hu, W.; Scholz, Y.; Yeligeti, M.; von Bremen, L.; Deng, Y. Downscaling ERA5 wind speed data: A machine learning approach considering topographic influences. \emph{Environ. Res. Lett.} \textbf{\boldmath{2023}}, \emph{18}, 094007. [\href{https://doi.org/10.1088/1748-9326/aceb0a}{CrossRef}]

\bibitem{B29-energies-3720435}
Chen, S.-T.; Yu, P.-S.; Tang, Y.-H. Statistical downscaling of daily precipitation using support vector machines and multivariate analysis. \emph{J. Hydrol.} \textbf{\boldmath{2010}}, \emph{385}, 13--22. [\href{https://doi.org/10.1016/j.jhydrol.2010.01.021}{CrossRef}]

\bibitem{B30-energies-3720435}
Pang, B.; Yue, J.; Zhao, G.; Xu, Z. Statistical Downscaling of Temperature with the Random Forest Model. \emph{Adv. Meteorol.} \textbf{\boldmath{2017}}, \emph{2017}, e7265178. [\href{https://doi.org/10.1155/2017/7265178}{CrossRef}]

\bibitem{B31-energies-3720435}
Sachindra, D.A.; Ahmed, K.; Rashid, M.M.; Shahid, S.; Perera, B.J.C. Statistical downscaling of precipitation using machine learning techniques. \emph{Atmos. Res.} \textbf{\boldmath{2018}}, \emph{212}, 240--258. [\href{https://doi.org/10.1016/j.atmosres.2018.05.022}{CrossRef}]

\bibitem{B32-energies-3720435}
Sekiyama, T.T.; Hayashi, S.; Kaneko, R.; Fukui, K. Surrogate Downscaling of Mesoscale Wind Fields Using Ensemble Superresolution Convolutional Neural Networks. \emph{Artif. Intell. Earth Syst.} \textbf{\boldmath{2023}}, \emph{2}, 230007. [\href{https://doi.org/10.1175/AIES-D-23-0007.1}{CrossRef}]

\bibitem{B33-energies-3720435}
Xu, R.; Chen, N.; Chen, Y.; Chen, Z. Downscaling and Projection of Multi-CMIP5 Precipitation Using Machine Learning Methods in the Upper Han River Basin. \emph{Adv. Meteorol.} \textbf{\boldmath{2020}}, \emph{2020}, e8680436. [\href{https://doi.org/10.1155/2020/8680436}{CrossRef}]

\bibitem{B34-energies-3720435}
Hobeichi, S.; Nishant, N.; Shao, Y.; Abramowitz, G.; Pitman, A.; Sherwood, S.; Bishop, C.; Green, S. Using Machine Learning to Cut the Cost of Dynamical Downscaling. \emph{Earth’s Future} \textbf{\boldmath{2023}}, \emph{11}, e2022EF003291. [\href{https://doi.org/10.1029/2022EF003291}{CrossRef}]

\bibitem{B35-energies-3720435}
Maclaurin, G.; Grue, N.; Lopez, A.; Heimiller, D.; Rossol, M.; Buster, G.; Williams, T. \emph{The Renewable Energy Potential (reV) Model: A Geospatial Platform for Technical Potential and Supply Curve Modeling}; NREL: Golden, CO, USA, 2021.

\bibitem{B36-energies-3720435}
Kim, J.; Lee, J.K.; Lee, K.M. Deeply-Recursive Convolutional Network for Image Super-Resolution. In Proceedings of the 2016 IEEE Conference on Computer Vision and Pattern Recognition (CVPR), Las Vegas, NV, USA, 27--30 June 2016; pp. 1637--1645. [\href{https://doi.org/10.1109/CVPR.2016.181}{CrossRef}]

\bibitem{B37-energies-3720435}
Tran, D.T.; Robinson, H.; Rasheed, A.; San, O.; Tabib, M.; Kvamsdal, T. GANs enabled super-resolution reconstruction of wind field. \emph{J. Phys. Conf. Ser.} \textbf{\boldmath{2020}}, \emph{1669}, 012029. [\href{https://doi.org/10.1088/1742-6596/1669/1/012029}{CrossRef}]

\bibitem{B38-energies-3720435}
Passarella, L.S.; Mahajan, S.; Pal, A.; Norman, M.R. Reconstructing High Resolution ESM Data Through a Novel Fast Super Resolution Convolutional Neural Network (FSRCNN). \emph{Geophys. Res. Lett.} \textbf{\boldmath{2022}}, \emph{49}, e2021GL097571. [\href{https://doi.org/10.1029/2021GL097571}{CrossRef}]

\bibitem{B39-energies-3720435}
Hu, X.; Naiel, M.A.; Wong, A.; Lamm, M.; Fieguth, P. RUNet: A Robust UNet Architecture for Image Super-Resolution. In Proceedings of the 2019 IEEE/CVF Conference on Computer Vision and Pattern Recognition Workshops (CVPRW), Long Beach, CA, USA, 16--17 June 2019. [\href{https://doi.org/10.1109/cvprw.2019.00073}{CrossRef}]

\bibitem{B40-energies-3720435}
Stengel, K.; Glaws, A.; Hettinger, D.; King, R.N. Adversarial super-resolution of climatological wind and solar data. \emph{Proc. Natl. Acad. Sci. USA} \textbf{\boldmath{2020}}, \emph{117}, 16805--16815. [\href{https://doi.org/10.1073/pnas.1918964117}{CrossRef}] [\href{https://www.ncbi.nlm.nih.gov/pubmed/32631993}{PubMed}]

\bibitem{B41-energies-3720435}
Chen, H.; Zhang, X.; Liu, Y.; Zeng, Q. Generative Adversarial Networks Capabilities for Super-Resolution Reconstruction of Weather Radar Echo Images. \emph{Atmosphere} \textbf{\boldmath{2019}}, \emph{10}, 555. [\href{https://doi.org/10.3390/atmos10090555}{CrossRef}]

\bibitem{B42-energies-3720435}
Jiang, Y.; Yang, K.; Shao, C.; Zhou, X.; Zhao, L.; Chen, Y.; Wu, H. A downscaling approach for constructing high-resolution precipitation dataset over the Tibetan Plateau from ERA5 reanalysis. \emph{Atmos. Res.} \textbf{\boldmath{2021}}, \emph{256}, 105574. [\href{https://doi.org/10.1016/j.atmosres.2021.105574}{CrossRef}]

\bibitem{B43-energies-3720435}
Ledig, C.; Theis, L.; Huszar, F.; Caballero, J.; Cunningham, A.; Acosta, A.; Aitken, A.; Tejani, A.; Totz, J.; Wang, Z.;~et~al. Photo-Realistic Single Image Super-Resolution Using a Generative Adversarial Network. In Proceedings of the 2017 IEEE Conference on Computer Vision and Pattern Recognition (CVPR), Honolulu, HI, USA, 21--26 July 2017; pp. 105--114. [\href{https://doi.org/10.1109/CVPR.2017.19}{CrossRef}]

\bibitem{B44-energies-3720435}
Yasuda, Y.; Onishi, R.; Matsuda, K. Super-resolution of three-dimensional temperature and velocity for building-resolving urban micrometeorology using physics-guided convolutional neural networks with image inpainting techniques. \emph{Build. Environ.} \textbf{\boldmath{2023}}, \emph{243}, 110613. [\href{https://doi.org/10.1016/j.buildenv.2023.110613}{CrossRef}]

\bibitem{B45-energies-3720435}
Rosencrans, D.; Benton, B.; Buster, G.; Glaws, A.; King, R.; Lundquist, J.; Gu, J.; Maclaurin, G. \emph{Wind Resource Data for Southeast Asia Using a Hybrid Numerical Weather Prediction with Machine Learning Super Resolution Approach, NREL/TP-5000-85481, 1984839, MainId:86254}; National Renewable Energy Laboratory (NREL): Golden, CO, USA, 2023.

\bibitem{B46-energies-3720435}
Bandura, R.; Romanishyn, A. Striving for Access, Security, and Sustainability: Ukraine’s Transition to a Modern and Decentralized Energy System. 2025. Available online: \href{https://www.csis.org/analysis/striving-access-security-and-sustainability}{\nolinkurl{https://www.csis.org/analysis/striving-access-security-and-sustainability}} (accessed on 3 July 2025).

\bibitem{B47-energies-3720435}
UNECE Renewable Energy Status Report 2022, Rana Adib, Executive Director, REN21\textbar{}UNECE. Available online: \href{https://unece.org/sed/documents/2022/11/presentations/unece-renewable-energy-status-report-2022-rana-adib-executive?utm_source=chatgpt.com}{\nolinkurl{https://unece.org/sed/documents/2022/11/presentations/unece-renewable-energy-status-report-2022-rana-adib-executive?utm\_source=chatgpt.com}} (accessed on 3 July 2025).

\bibitem{B48-energies-3720435}
Prengaman, P. Ukraine Has Seen Success in Building Clean Energy, Which Is Harder for Russia to Destroy, AP News. Available online: \href{https://apnews.com/article/ukraine-clean-renewable-energy-russian-bombing-distributed-1f226213742cc057f9f65208167e6f38}{\nolinkurl{https://apnews.com/article/ukraine-clean-renewable-energy-russian-bombing-distributed-1f226213742cc057f9f65208167e6f38}} (accessed on 3 July 2025).

\bibitem{B49-energies-3720435}
Hersbach, H.; Bell, B.; Berrisford, P.; Hirahara, S.; Horányi, A.; Muñoz-Sabater, J.; Nicolas, J.; Peubey, C.; Radu, R.; Schepers, D.;~et~al. The ERA5 global reanalysis. \emph{Quart J. R. Meteoro. Soc.} \textbf{\boldmath{2020}}, \emph{146}, 1999--2049. [\href{https://doi.org/10.1002/qj.3803}{CrossRef}]

\bibitem{B50-energies-3720435}
Kalnay, E. \emph{Atmospheric Modeling, Data Assimilation and Predictability}; Cambridge University Press: Cambridge, UK, 2003.

\bibitem{B51-energies-3720435}
Sheridan, L.M.; Phillips, C.; Orrell, A.C.; Berg, L.K.; Tinnesand, H.; Rai, R.K.; Zisman, S.; Duplyakin, D.; Flaherty, J.E. Validation of wind resource and energy production simulations for small wind turbines in the United States. \emph{Wind Energy Sci.} \textbf{\boldmath{2022}}, \emph{7}, 659--676. [\href{https://doi.org/10.5194/wes-7-659-2022}{CrossRef}]

\bibitem{B52-energies-3720435}
Estima, J.; Fichaux, N.; Menard, L.; Ghedira, H. The global solar and wind atlas: A unique global spatial data infrastructure for all renewable energy. In Proceedings of the 1st ACM SIGSPATIAL International Workshop on MapInteraction, in MapInteract ’13, New York, NY, USA, 5 November 2013; Association for Computing Machinery: New York, NY, USA, 2013; pp. 36--39. [\href{https://doi.org/10.1145/2534931.2534933}{CrossRef}]

\bibitem{B53-energies-3720435}
NOAA NCEP Meteorological Assimilation Data Ingest System (MADIS). Available online: \href{https://madis.ncep.noaa.gov/}{\nolinkurl{https://madis.ncep.noaa.gov/}} (accessed on 6 December 2023).

\bibitem{B54-energies-3720435}
Wilczak, J.M.; Stoelinga, M.; Berg, L.K.; Sharp, J.; Draxl, C.; McCaffrey, K.; Banta, R.M.; Bianco, L.; Djalalova, I.; Lundquist, J.K.;~et~al. The Second Wind Forecast Improvement Project (WFIP2): Observational Field Campaign. \emph{Bull. Am. Meteorol. Soc.} \textbf{\boldmath{2019}}, \emph{100}, 1701--1723. [\href{https://doi.org/10.1175/BAMS-D-18-0035.1}{CrossRef}]

\bibitem{B55-energies-3720435}
Complete ERA5 Global Atmospheric Reanalysis. Available online: \href{https://cds.climate.copernicus.eu/datasets/reanalysis-era5-complete?tab=overview}{\nolinkurl{https://cds.climate.copernicus.eu/datasets/reanalysis-era5-complete?tab=overview}} (accessed on 3 July 2025).

\bibitem{B56-energies-3720435}
Dee, D.P.; Uppala, S.M.; Simmons, A.J.; Berrisford, P.; Poli, P.; Kobayashi, S.; Andrae, U.; Balmaseda, M.A.; Balsamo, G.; Bauer, P.;~et~al. The ERA-Interim reanalysis: Configuration and performance of the data assimilation system. \emph{Q. J. R. Meteorol. Soc.} \textbf{\boldmath{2011}}, \emph{137}, 553--597. [\href{https://doi.org/10.1002/qj.828}{CrossRef}]

\bibitem{B57-energies-3720435}
USGS EROS Archive-Digital Elevation-Global 30 Arc-Second Elevation (GTOPO30)\textbar{}U.S. Geological Survey. Available online: \href{https://www.usgs.gov/centers/eros/science/usgs-eros-archive-digital-elevation-global-30-arc-second-elevation-gtopo30}{\nolinkurl{https://www.usgs.gov/centers/eros/science/usgs-eros-archive-digital-elevation-global-30-arc-second-elevation-gtopo30}} (accessed on 7 July 2025).

\bibitem{B58-energies-3720435}
Benton, B.; Buster, G.; Glaws, A.; King, R. \emph{sup3r (Super Resolution for Renewable Resource Data)}; National Renewable Energy Lab. (NREL): Golden, CO, USA, 2022. Available online: \href{https://zenodo.org/records/10402581}{\nolinkurl{https://zenodo.org/records/10402581}} (accessed on 14 July 2025).

\bibitem{B59-energies-3720435}
Wilczak, J.M.; Akish, E.; Capotondi, A.; Compo, G.P. Evaluation and Bias Correction of the ERA5 Reanalysis over the United States for Wind and Solar Energy Applications. \emph{Energies} \textbf{\boldmath{2024}}, \emph{17}, 1667. [\href{https://doi.org/10.3390/en17071667}{CrossRef}]

\bibitem{B60-energies-3720435}
Millstein, D.; Jeong, S.; Ancell, A.; Wiser, R. A database of hourly wind speed and modeled generation for US wind plants based on three meteorological models. \emph{Sci. Data} \textbf{\boldmath{2023}}, \emph{10}, 883. [\href{https://doi.org/10.1038/s41597-023-02804-w}{CrossRef}] [\href{https://www.ncbi.nlm.nih.gov/pubmed/38065988}{PubMed}]

\bibitem{B61-energies-3720435}
Potisomporn, P.; Adcock, T.A.A.; Vogel, C.R. Evaluating ERA5 reanalysis predictions of low wind speed events around the UK. \emph{Energy Rep.} \textbf{\boldmath{2023}}, \emph{10}, 4781--4790. [\href{https://doi.org/10.1016/j.egyr.2023.11.035}{CrossRef}]

\bibitem{B62-energies-3720435}
Bouall{\fontencoding{T5}\selectfont{\`e}}gue, Z.B.; Cooper, F.; Chantry, M.; Düben, P.; Bechtold, P.; Sandu, I. Statistical Modeling of 2-m Temperature and 10-m Wind Speed Forecast Errors. \emph{Mon. Weather. Rev.} \textbf{\boldmath{2023}}, \emph{151}, 897--911. [\href{https://doi.org/10.1175/MWR-D-22-0107.1}{CrossRef}]

\end{thebibliography}
\end{document}